\documentclass[a4paper,3p]{elsarticle}
\journal{}
\usepackage{lineno}
\modulolinenumbers[5]
\usepackage{hyperref}
\usepackage{graphicx,caption}
\usepackage{cmap}
\usepackage[T1]{fontenc}
\usepackage[utf8]{inputenc}
\usepackage[english]{babel}
\usepackage{textcomp}
\usepackage{xcolor}
\usepackage{amsmath}
\usepackage{amssymb}
\usepackage{bm}
\usepackage[normalem]{ulem}
\usepackage{physics}
\usepackage{multirow}
\usepackage{hhline}
\usepackage{subfigure}
\usepackage{comment}
\usepackage{float}
\usepackage{multicol}
\usepackage{blindtext}

\begin{document}
	
\begin{frontmatter}
\title{A family of conservative axisymmetric contact SPH schemes for impact engineering applications}
\author{G.~D.~Rublev$^{1,2}$}
\ead{rublev\_gd\_97@vk.com} 
\address{$^{1)}$ Dukhov  Research Institute of Automatics, Russia, Moscow 127030, Sushchevskaya St., 22}
\address{$^{2)}$ Joint Institute for High Temperatures of RAS, Moscow, 125412, Russia}
\begin{abstract}\noindent
A family of conservative schemes for the axisymmetric contact smoothed particle hydrodynamics (CSPH) method, which ensure the accuracy and stability in modeling of complex multi-material flows of compressible media, is introduced. Among these schemes, the most convenient ones are considered. Simulations with the proposed schemes may be also improved by embedding of MUSCL reconstruction into a numerical scheme, or by correcting the kernel gradient as was proposed earlier for the Cartesian case. 

Verification of the proposed method is performed on several test problems: Sod's cylindrical test, Taylor bar test, and Sedov's point explosion. The conservative properties of the scheme are demonstrated. Finally, a set of simulations on air shock wave weakening by a breakaway sand barrier is performed and compared to experimental results. 
\end{abstract}
\end{frontmatter}

\section*{Introduction}

Smoothed Particles Hydrodynamics (SPH) method, which was initially developed for astrophysical applications~\cite{Lucy1977,Monaghan:MNRAS:1977}, became a popular tool nowadays for the compressible media modeling. Its meshless nature allows to overcome some restrictions of conventional mesh-based Euler and Lagrange approaches, especially when modeling free surface or multi-phase flows~\cite{Colagrossi:RPP:2025} and large strain phenomena in elastic-plastic media~\cite{Mehra2012,dyachkov2018explicit,Ikkurthi2023}. 

It is worth noting, that the stability of earlier SPH schemes for compressible flows simulations was reached by introducing the artificial viscosity~\cite{Monaghan:JCP:1983,Monaghan1988}. This approach usually requires the artificial viscosity tuning for various problems, and the choice of suitable parameters is not always trivial. As an alternative, following with the development of mesh-based Godunov type methods, the Riemann solution was also implemented in SPH~\cite{Parshikov2000,Parshikov:JCP:2002,Inutsuka2002}. The momentum and energy fluxes in of the Godunov-like contact SPH (CSPH) method~\cite{Parshikov:JCP:2002} are computed using the solution of the Riemann problem based on sound wave approximation at the contact between the base particle and each of the particles in its environment. In addition, this approach was provided with extensions for modeling of elastoplastic media and heat conduction. 

Main disadvantage of the CSPH method is high numerical viscosity, which causes smoothing of shock wave fronts and unphysical damping of oscillations in media. As a result, modeling of flows with realistic viscosity was almost impossible. The approaches for reducing numerical dissipation were developed~\cite{Zhang2017}, which indirectly reduce the numerical viscosity. Recently, the correction of CSPH equations was provided~\cite{Parshikov2023} which utilizes the compensating viscosity tensor. As a result, CSPH method with the incorporated Riemann problem solution was also extended to model viscous compressible flows.

The majority of axisymmetric SPH method formulations are based on the artificial viscosity approach. I was first proposed in 1985 by Coleman and Bicknell~\cite{Coleman1985}. In 1993 Petschek~\cite{Petschek1993} proposed an axisymmetric version of the SPH method, which was obtained by a direct transition from the three-dimensional Cartesian case with Gaussian kernel to the two-dimensional axisymmetric case by angle integration. In subsequent years, numerous alternative variants of the axisymmetric SPH method were proposed, wherein artificial viscosity was utilized (see, for instance, \cite{OMANG2006391,Li2020}). The axisymmetric SPH method with the Riemann problem solution at interparticle contacts was first proposed by Parshikov in Ref.~\cite{Parshikov2000}, but the derived SPH scheme was not conservative. In 2022, Xiang-Li Fang, Andrea Colagrossi and others~\cite{Fang2022} proposed to introduce MUSCL reconstruction into A.N.~Parshikov's axisymmetric method~\cite{Parshikov2000} to model underwater explosions.

The CSPH method~\cite{Parshikov:JCP:2002} and its modern extensions~\cite{XiaoyanHu2016,Meng2021,Ren2023,Rublev2024} are nowadays an advanced framework for modeling compressible media using meshless approach. However, three-dimensional modeling in the Cartesian coordinates may require large number of particles and can be a time-consuming process. For a number of practically important problems, two-dimensional axisymmetric modeling can be used, which can significantly speed up the simulation, reduce the computational load, and enhance the spatial resolution: for example, in modeling obstacle penetration~\cite{Parshikov2000, Johnson1993}, for modeling axisymmetric fluid flows, shell compression problems, bubbles in fluid~\cite{Li2020}, modeling explosions in the atmosphere, underwater and underground. The axisymmetric CSPH method can also be useful in modeling the effects of high-power laser radiation on matter~\cite{grigoryev2018expansion}. Thus, the development of axisymmetric SPH contact methods is considered to be a subject of significant importance by the author.

In the proposed work, the axisymmetric version of the CSPH method is advanced by providing conservative equations, including the use of MUSCL reconstruction~\cite{XiaoyanHu2016,Fang2022} and Total Kernel Correction (TKC) technique~\cite{Rublev2024} to reduce numerical diffusion. A significant number of features of the axisymmetric SPH method have not been the subject of comprehensive study, and the present paper addresses this lacuna. To develop a conservative contact axisymmetric SPH method, it is necessary to choose such a conservative scheme of the conventional method, where, without violating the conservativity of the scheme, it is possible to replace the pressure and velocity half-sums of the base particle and each neighboring particle by the corresponding contact values. To the best of the author's knowledge, there are no suitable schemes among the available ones~\cite{Petschek1993,Broo2003,Omang2006,Neves2015}.

To demonstrate the advantages of the proposed schemes, modeling of a number of Sod's cylindrical test, Taylor bar test, and Sedov's point explosion has been carried out. However, impact engineering applications may include complex setups with various materials and models. To validate the advanced axisymmetric CSPH method, the simulation of shock wave weakening by a breakaway sand barrier is carried out. Employing a low-strength breakaway barrier is often used to mitigate the shock wave impact on a protected object. The velocity difference that occurs as a result of the explosion leads to the transfer of momentum from the low-density ``light'' medium (comprising air and explosion products) to the fine, low-velocity ``heavy'' fraction (sand) that is formed during the fragmentation of the barrier. The shock wave and the material flow behind it weaken to become safe for the protected object, however the sand fraction accelerates, but not exceeding the safety limits.

Computer simulation of such an experiment is rather challenging: the mesh-based Lagrange methods would require dramatic mesh rearrangement, while the use of Euler methods necessitates the employment of the equations of two-velocity hydrodynamics. This is due to the fact that the fragments of the destroyed barrier (sand grains) acquire velocity in the overtaking explosion products due to the force and viscous interaction with the latter. In this instance, the sand and explosion products should be modeled using multi-phase approaches similar to Ref.~\cite{BN}. Furthermore, it is necessary to develop a model of interaction of these continuums to construct exchange terms in the right-hand sides of the equations of multi-phase hydrodynamics. In this case, the correctness of the solution of the problem depends solely on  the assumptions adopted in the construction of the exchange terms.

Thus it appears that the meshless smoothed particle hydrodynamics (SPH) method is a more suitable approach. In this case, each of the contacting media (explosion products, sand, air medium) occupies its own region of space and is described by its own variety of Lagrangian particles. Consequently, the single-velocity approximation is valid, and the velocity disequilibrium between the explosion products and sand fragments is appears naturally. The modeling results obtained with the proposed axisymmetric CSPH scheme demonstrate a close alignment with the experimental data.

The paper structured as follows: Section~\ref{sec:axisymmetric-csph} provides the derivation of the equations of the proposed method. Section~\ref{BC_and_particle_adoptation} presents additional numerical algorithms that violate conservativity, but are required to improve the quality of simulation results (boundary condition near the axis of symmetry and particle splitting/merging algorithms).  Section~\ref{Numerical-examples} contains numerical tests for the proposed method. Section~\ref{SW-weakening} displays validation results on shock wave weakening by a breakaway sand barrier experimental data. Description of the adopted TKC (Total Kernel Correction) procedure used is provided in the appendix.


\section{Axisymmetric contact SPH method}
\label{sec:axisymmetric-csph}

Throughout this paper tensor notations instead of index notations are used. Vectors and tensors are denoted in bold. The definitions of tensor operations used further in terms of indices (assuming summation by repeated indices) are as follows:
$$\mathbf{A} \cdot \mathbf{B} = A_{\alpha\ldots\beta}B_{\beta\ldots\gamma},$$
$$\mathbf{A} : \mathbf{B} = A_{\alpha\ldots\beta\gamma}B_{\beta\gamma\ldots\delta},$$
$$\mathbf{A} \otimes \mathbf{B} = A_{\alpha\ldots\beta}B_{\gamma\ldots\delta}.$$

\subsection{Governing axisymmetric equations}

The model of compressible media is formulated using the equations expressing the laws of conservation of mass, momentum, and energy (in differential form) are used in the axisymmetric case. The continuity equation is written in the form of the equation for the volumetric strain:
\begin{equation}\label{eq:continuity-cyl}
	\frac{d\varepsilon}{dt}=-\frac{1}{\rho}\frac{d\rho}{dt} = \frac{1}{r}\frac{\partial}{\partial r}\left(rU^r\right) + \frac{\partial U^z}{\partial z} = \boldsymbol{\nabla}\cdot\mathbf{U} + \frac{U^r}{r},
\end{equation}
where
$$\boldsymbol{\nabla}\cdot\mathbf{U} =\frac{\partial U^r}{\partial r} + \frac{\partial U^z}{\partial z},$$
$\boldsymbol{\nabla} = (\partial/\partial r, \partial/\partial z)^\text{T}$, $\mathbf{U} = (U^r, U^z)^\text{T}$ is the velocity, $\rho$ is the density.
The equation of motion is:
\begin{equation}\label{eq:momentum-cyl}
	\frac{d\mathbf{U}}{dt}=-\frac{1}{\rho}\left(\frac{\partial P}{\partial r}\mathbf{e}^r + \frac{\partial P}{\partial z}\mathbf{e}^z\right) = -\frac{1}{\rho}\boldsymbol{\nabla} P,
\end{equation}
where $P$ is pressure, $\mathbf{e}^{\alpha}$ are unit vectors of the cylindrical coordinate system $r$-$z$.
The evolution of the internal energy using the notations introduced above is expressed as:
\begin{equation}\label{eq:energy-cyl}
	\frac{de}{dt}=-\frac{P}{\rho}\left(\frac{1}{r}\frac{\partial}{\partial r}\left(rU^r\right) + \frac{\partial U^z}{\partial z}\right) =  -\frac{P}{\rho}\boldsymbol{\nabla}\cdot\mathbf{U} - \frac{PU^r}{\rho r},
\end{equation}
where $e$ is the specific internal energy.

The system of equations (\ref{eq:continuity-cyl})--(\ref{eq:energy-cyl}) is closed by the equation of state
$$P = P(\rho,e).$$

\subsection{Smoothed particles approximation}

A scalar function $F$ defined in the half-plane $r$-$z$ ($r>0$) may be approximated via the smoothing kernel $W$ at the point $\mathbf{r} = (r,\, z)^T$ as:
\begin{equation}\label{eq:scalar-kernel-approximation-axisymmetric}
	F(\mathbf{r}) = \int dr'\int dz' F(\mathbf{r}')W(\left|\left|\mathbf{r}' - \mathbf{r}\right|\right|/h;h) + O(h) =
	\\
	\int ds' F(\mathbf{r}')W(\left|\left|\mathbf{r}' - \mathbf{r}\right|\right|/h;h) + O(h),
\end{equation}
where $ds = dr\cdot dz$ is the area element. The smoothing kernel $W\left(q;h\right)$ is a function of $q = \left|\left|\mathbf{r} - \mathbf{r'}\right|\right|/h$ depending on the smoothing length $h$ as a parameter.

Let $m_a$ be the mass of a toroidal material element (a particle) of the radius $r_a$ with the cross-sectional area of
$$S_a = \frac{m_a}{2 \pi r_a \rho_a}.$$
Then the particle size $D_a$ is $S_a^{1/d},$ where $d$ is the dimension of the problem ($1$ or $2$).

The SPH approximation of the scalar function $F$ value at the point $a$ in the axisymmetric case is:
\begin{equation}\label{eq:scalar-sph-approximation-axisymmetric}
	\left<F\right>_a = \sum\limits_{b}F_b\frac{m_b}{2\pi r_b\rho_b}W_{ab}.
\end{equation}
Derivative of the equation (\ref{eq:scalar-sph-approximation-axisymmetric}) provide an estimate for the gradient of the function $F$ at the point $a$:
\begin{equation}\label{eq:scalar-sph-approximation-grad-axisymmetric}
	\left<\boldsymbol{\nabla} F\right>_a = \sum\limits_{b}F_b\frac{m_b}{2\pi r_b\rho_b}\boldsymbol{\nabla}_aW_{ab},
\end{equation}
where
$$\boldsymbol{\nabla}_a W_{ab} = -W'_{ab}\frac{\mathbf{e}^R_{ba}}{h_{ab}}, \,\,\,\, W'(q;h) \equiv \frac{dW}{dq}(q;h),$$ 
$\mathbf{e}^R_{ba} = (\mathbf{r}_{b} - \mathbf{r}_{a})/||\mathbf{r}_{b} - \mathbf{r}_{a}||$, $h_{ab} = \theta (D_a + D_b)$ is the smoothing length, $\theta$ is the smoothing length parameter.

Similarly for the vector field $\mathbf{F}$ one can obtain:
\begin{equation}\label{eq:vector-sph-approximation-div-axisymmetric}
	\left<\boldsymbol{\nabla}\cdot\mathbf{F}\right>_a = \sum\limits_{b}\frac{m_b}{2\pi r_b\rho_b}\mathbf{F}_b\cdot\boldsymbol{\nabla}_aW_{ab},
\end{equation}
\begin{equation}\label{eq:vector-sph-approximation-otimes-axisymmetric}
	\left<\boldsymbol{\nabla}\otimes\mathbf{F}\right>_a = \sum\limits_{b}\frac{m_b}{2\pi r_b\rho_b}\boldsymbol{\nabla}_aW_{ab}\otimes \mathbf{F}_b,
\end{equation}
where
\begin{equation*}
	\boldsymbol{\nabla}\otimes\mathbf{F} = 
	\begin{pmatrix}
		\frac{\partial F^r}{\partial r} & \frac{\partial F^z}{\partial r} \\
		\frac{\partial F^r}{\partial z} & \frac{\partial F^z}{\partial z}
	\end{pmatrix}.
\end{equation*}
Further, for brevity, the angle brackets are omitted when writing SPH approximations.


The concept of the contact CSPH method consists in introducing into the SPH approximation the velocities and stresses (pressures) on the contact surface obtained from the solution of the Riemann problem instead of the average values of velocities and stresses (pressures) between pairs of interacting particles, as it was proposed in the traditional SPH method. Therefore, it is required to obtain the basic equations in a form where the half-sums of velocities and stresses (pressures) can be easily replaced by the corresponding contact values obtained from the solution of the Riemann problem.

\subsection{A family of CSPH schemes preserving conservativity}

Let $f(r;A)$ be a smooth function of the radial coordinate, depending on $A$ as a parameter. The pressure gradient may be approximated using the following identity:
\begin{equation}\label{eq:tozhdestvo-1}
	\boldsymbol{\nabla} P = \frac{\boldsymbol{\nabla} (f(r;A)P) - P\boldsymbol{\nabla} f(r;A)}{f(r;A)} =
	\frac{\boldsymbol{\nabla} (f(r;A)P) + P\boldsymbol{\nabla} f(r;A)}{f(r;A)} - 2P\frac{f'(r;A)}{f(r;A)}\mathbf{e}^r.
\end{equation}

The SPH approximation of the equation of motion is obtained via (\ref{eq:scalar-sph-approximation-grad-axisymmetric}) and (\ref{eq:tozhdestvo-1}) as:
\begin{equation}\label{eq:axisym-general-momentum}
	\frac{d\mathbf{U}_a}{dt} = -\sum\limits_{b}\frac{m_b}{2\pi \rho_b\rho_ar_b}\frac{f(r_b;A)}{f(r_a;A)}(P_a + P_b)\boldsymbol{\nabla}_aW_{ab} + 2\frac{P_af'(r_a;A)}{\rho_a f(r_a;A)}\mathbf{e}^r.
\end{equation}

The divergence of velocity may be approximated using the following identity:
\begin{equation}\label{eq:tozhdestvo-2}
	\boldsymbol{\nabla}\cdot \mathbf{U} = \frac{\boldsymbol{\nabla} \cdot(f(r;A)\mathbf{U}) - \mathbf{U}\cdot\boldsymbol{\nabla} f(r;A)}{f(r;A)}.
\end{equation}
Then using (\ref{eq:vector-sph-approximation-div-axisymmetric}) and (\ref{eq:tozhdestvo-2}) one can obtain the continuity equation in SPH form as follows:
\begin{equation}\label{eq:axisym-general-continuity}
	\frac{d\varepsilon_a}{dt} = \sum\limits_{b}\frac{m_b}{2\pi \rho_b r_b}\frac{f(r_b;A)}{f(r_a;A)}(\mathbf{U}_b - \mathbf{U}_a)\cdot\boldsymbol{\nabla}_aW_{ab} + \frac{U^r_a}{r_a}.
\end{equation}

The equation for internal energy evolution may be written as:
\begin{equation*}
	\frac{de_a}{dt} = -\sum\limits_{b}\frac{m_b}{2\pi \rho_b\rho_a r_b}\frac{f(r_b;A)}{f(r_a;A)}\frac{P_a+P_b}{2}(\mathbf{U}_b - \mathbf{U}_a)\cdot\boldsymbol{\nabla}_aW_{ab} - \frac{P_aU^r_a}{\rho_a r_a}.
\end{equation*}
Then the equation for the total energy takes the form:
\begin{multline}\label{eq:axisym-general-total-energy}
	\frac{dE_a}{dt} = \frac{de_a}{dt} + \mathbf{U}_a\cdot\frac{d\mathbf{U}_a}{dt} =
	-\sum\limits_{b}\frac{m_b}{2\pi \rho_b\rho_a r_b}\frac{f(r_b;A)}{f(r_a;A)}\frac{P_a+P_b}{2}(\mathbf{U}_b + \mathbf{U}_a)\cdot\boldsymbol{\nabla}_aW_{ab} +
	\\
	\left(\frac{2f'(r_a;A)}{f(r_a;A)}-\frac{1}{r_a}\right)\frac{P_aU^r_a}{\rho_a}.
\end{multline}

Further the parameter $A$ in equations (\ref{eq:axisym-general-momentum}), (\ref{eq:axisym-general-continuity}), (\ref{eq:axisym-general-total-energy}) is assumed to be equal to $r_a$. As can be seen from equation (\ref{eq:axisym-general-momentum}), in order for the total momentum to be conserved, it must be required that the following holds
\\
\textbf{Condition 1.} $$\frac{f(r_b;r_a)}{r_bf(r_a;r_a)} = \frac{f(r_a;r_b)}{r_af(r_b;r_b)},\,\,\,\, \forall r_a,\,r_b.$$
It follows from equation (\ref{eq:axisym-general-total-energy}) that for the conservativity of the total energy equation one must require the following condition be satisfied:
\\
\textbf{Condition 2.} $$2\frac{f'(r_a;r_a)}{f(r_a;r_a)} = \frac{1}{r_a}.$$

Without restricting generality, let's assume that $f(r_a;r_a)=1$ for any $r_a$. Let us consider the function $f(r;A)$ in the form:
\begin{equation}\label{eq:g-function}
	f(r;A) = \frac{1}{A}g(r,A),
\end{equation}	
where $g$ is a symmetric function of its arguments, that is, $g(r,A) = g(A,r),$ with $g(r_a, r_a) = r_a.$ It follows from Condition 2 that $g'(r_a, r_a)=1/2,$ where it is implied that the differentiation is performed on the first argument.

After introducing the solution of the Riemann problem into equations (\ref{eq:axisym-general-momentum}), (\ref{eq:axisym-general-continuity}), (\ref{eq:axisym-general-total-energy}), the general form of the equations of the specified family of schemes of the axisymmetric CSPH method can be obtained:
\begin{equation}\label{eq:axisym-contact-general-continuity}
	\frac{d\varepsilon_a}{dt} = -\sum\limits_{b}\frac{m_b}{2\pi \rho_b r_bh_{ab}}f(r_b;r_a)2(U_{ab}^{R*} - U_a^R)W'_{ab} + \frac{U^r_a}{r_a}.
\end{equation}
\begin{equation}\label{eq:axisym-contact-general-momentum}
	\frac{d\mathbf{U}_a}{dt} = -\sum\limits_{b}\frac{m_b}{2\pi \rho_b\rho_ar_b}f(r_b;r_a)2P_{ab}^*\boldsymbol{\nabla}_aW_{ab} + \frac{P_a}{\rho_a r_a}\mathbf{e}^r.
\end{equation}
\begin{equation}\label{eq:axisym-contact-general-total-energy}
	\frac{dE_a}{dt} = 
	\sum\limits_{b}\frac{m_b}{2\pi \rho_b\rho_a r_bh_{ab}}f(r_b;r_a)2P_{ab}^*U_{ab}^{R*}W'_{ab},
\end{equation}
where
\begin{equation}\label{RazpadP}
	P^{*}_{ab} = \frac{P_rZ_l + P_lZ_r - Z_lZ_r\left( U_{r}^R - U_{l}^R \right)}{Z_l  + Z_r},
\end{equation}
\begin{equation}
	\label{RazpadU}
	U^{R*}_{ab} = \frac{U_l^{R}Z_l + U_r^{R}Z_r - P_r + P_l}{Z_l  + Z_r},
\end{equation}
$Z = \rho C$ is the acoustic impedance of the material, $C$ is the sound speed.

Conservative schemes for the contact axisymmetric CSPH method can be obtained by selecting a function $g$ satisfying the above conditions.
Table \ref{table:axisym-schemes-examples} provides examples of the function $g$ choices and the corresponding schemes. Depending on the combination of the radial coordinates $r_a$ and $r_b$ in the right-hand sides of the equations, i.e., on the factors $2/{\sqrt{r_ar_b}}$, $ 4/(r_a+r_b)$, and $ (r_a+r_b)/r_ar_b$, will hereafter be referred to the schemes for the contact axisymmetric CSPH method as ``geometric mean'', ``arithmetic mean'', and ``harmonic mean'', respectively. In the following, the ``harmonic mean'' scheme is used, since the function $f(r,r_a)$ corresponding to this scheme is continuously differentiable everywhere if $r_a \neq 0$.

\begin{table}
	\center{
		\caption{\label{table:axisym-schemes-examples} Examples of conservative axisymmetric schemes from the family (\ref{eq:axisym-contact-general-continuity}-\ref{eq:axisym-contact-general-total-energy}).}
		\tabcolsep10pt
		\renewcommand{\arraystretch}{1.8}
		\begin{tabular}{c|c|c}
			\hline
			\Large $g(r, r_a)$		& Scheme  & Name      \\
			\hline
			&  \Large $\frac{d\varepsilon_a}{dt} = -\sum\limits_{b}\frac{m_b}{2\pi \rho_b h_{ab}}\frac{2}{\sqrt{r_ar_b}}(U_{ab}^{R*} - U_a^R)W'_{ab} + \frac{U^r_a}{r_a}$  &
			\\
			\Large $\sqrt{rr_a}$   & \Large $\frac{d\mathbf{U}_a}{dt} = -\sum\limits_{b}\frac{m_b}{2\pi \rho_b\rho_a}\frac{2}{\sqrt{r_ar_b}}P_{ab}^*\boldsymbol{\nabla}_aW_{ab} + \frac{P_a}{\rho_a r_a}\mathbf{e}^r$  & $\begin{matrix}\text{``Geometric} \\ \text{mean''}\end{matrix}$ \\
			& \Large $\frac{dE_a}{dt} = 
			\sum\limits_{b}\frac{m_b}{2\pi \rho_b\rho_a  h_{ab}}\frac{2}{\sqrt{r_ar_b}}P_{ab}^*U_{ab}^{R*}W'_{ab}$ & \\
			\hline
			&   \Large $\frac{d\varepsilon_a}{dt} = -\sum\limits_{b}\frac{m_b}{2\pi \rho_b h_{ab}}\frac{4}{r_a+r_b}(U_{ab}^{R*} - U_a^R)W'_{ab} + \frac{U^r_a}{r_a}$ &
			\\
			\Large $\frac{2r_ar}{r+r_a}$   & \Large $\frac{d\mathbf{U}_a}{dt} = -\sum\limits_{b}\frac{m_b}{2\pi\rho_b \rho_a}\frac{4}{r_a+r_b}P_{ab}^*\boldsymbol{\nabla}_aW_{ab} + \frac{P_a}{\rho_a r_a}\mathbf{e}^r$  &  $\begin{matrix}\text{``Arithmetic} \\ \text{mean''}\end{matrix}$\\
			& \Large $\frac{dE_a}{dt} = 
			\sum\limits_{b}\frac{m_b}{2\pi \rho_b\rho_a  h_{ab}}\frac{4}{r_a+r_b}P_{ab}^*U_{ab}^{R*}W'_{ab}$ & \\
			\hline			
			&   \Large $\frac{d\varepsilon_a}{dt} = -\sum\limits_{b}\frac{m_b}{2\pi \rho_b h_{ab}}\frac{(r_a+r_b)}{r_ar_b}(U_{ab}^{R*} - U_a^R)W'_{ab} + \frac{U^r_a}{r_a}$ & 
			\\
			\Large $\frac{r_a+r}{2}$   & \Large $\frac{d\mathbf{U}_a}{dt} = -\sum\limits_{b}\frac{m_b}{2\pi\rho_b \rho_a}\frac{(r_a+r_b)}{r_ar_b}P_{ab}^*\boldsymbol{\nabla}_aW_{ab} + \frac{P_a}{\rho_a r_a}\mathbf{e}^r$  &  $\begin{matrix}\text{``Harmonic} \\ \text{mean''}\end{matrix}$\\
			& \Large $\frac{dE_a}{dt} = 
			\sum\limits_{b}\frac{m_b}{2\pi \rho_b\rho_ah_{ab}}\frac{(r_a+r_b)}{r_ar_b}P_{ab}^*U_{ab}^{R*}W'_{ab}$ & \\
			\hline	
		\end{tabular}
		\tabcolsep15pt
	}
\end{table}

It should be noted that the constructed schemes of the axisymmetric CSPH method provide non-decreasing entropy. Lets show it on the example of the ``harmonic mean'' scheme. According to the first principle of thermodynamics:

\begin{equation}\label{Tds}
	T_a\frac{ds_a}{dt} = -\frac{P_a}{\rho_a^2}\frac{d\rho_a}{dt} + \frac{de_a}{dt} = \frac{P_a}{\rho_a}\frac{d\varepsilon_a}{dt} + \frac{de_a}{dt},
\end{equation}
where $T$ is temperature, $s$ is specific entropy. From the equations of motion and energy one can obtain the equation for internal energy:
\begin{equation}\label{eq:csph-internal-energy-axisymmetric-harmonic}
	\begin{gathered}
		\frac{de_a}{dt} = \frac{dE_a}{dt} - \mathbf{U}_a\cdot\frac{d\mathbf{U}_a}{dt} = 
		\\
		-\sum\limits_{b}\frac{m_b}{2\pi\rho_a\rho_b}\frac{(r_a+r_b)}{r_ar_b}P_{ab}^*(\mathbf{U}_{ab}^{*} - \mathbf{U}_a)\cdot\boldsymbol{\nabla}_aW_{ab} - \frac{P_aU^r_a}{\rho_ar_a} = 
		\\
		\sum\limits_{b}\frac{m_b}{2\pi\rho_a\rho_b}\frac{(r_a+r_b)}{r_ar_b}P_{ab}^*(U_{ab}^{R*} - U_a^R)\frac{W'_{ab}}{h_{ab}} - \frac{P_aU^r_a}{\rho_ar_a}.
	\end{gathered}
\end{equation}
After substituting the continuity equation and the internal energy equation (\ref{eq:csph-internal-energy-axisymmetric-harmonic}) into (\ref{Tds}):
\begin{equation}\label{Tds-SPH}
	T_a\frac{ds_a}{dt} = \sum\limits_{b}\frac{m_b}{2\pi\rho_a\rho_b}\frac{(r_a+r_b)}{r_ar_b}(P_{ab}^* - P_a)(U_{ab}^{R*} - U_a^R)\frac{W'_{ab}}{h_{ab}}.
\end{equation}
Using explicit expressions for the contact values in the acoustic approximation, one can obtain:
\begin{equation*}
	T_a\frac{ds_a}{dt} = -\sum\limits_{b}\frac{m_b}{2\pi\rho_a\rho_b}\frac{(r_a+r_b)}{r_ar_b}\frac{Z_a}{(Z_a+Z_b)^2}\left[(U^R_b - U^R_a)Z_b - (P_b - P_a)\right]^2\frac{W'_{ab}}{h_{ab}} \geq 0,
\end{equation*}
which ensures non-decreasing entropy.

The equations of the CSPH method for an elastic medium may be obtained as follows. In general, the stress tensor in an elastic medium consists of two parts:
$$\boldsymbol{\sigma} = -P\mathbf{I} + \mathbf{S},$$
where $\mathbf{I}$ is the unit tensor, $\mathbf{S}$ is the elastic stress deviator tensor. 
In this case, the equation of motion takes the form:
\begin{equation}\label{eq:momentum-strength-cyl}
	\frac{d\mathbf{U}}{dt} = \frac{d}{dt}
	\begin{pmatrix}
		U^r
		\\
		U^z
	\end{pmatrix}
	=\frac{1}{\rho}
	\begin{pmatrix}
		\frac{\partial \sigma^{rr}}{\partial r} + \frac{\partial \sigma^{zr}}{\partial z} + \frac{\sigma^{rr} -  \sigma^{\theta\theta}}{r}
		\\
		\frac{\partial \sigma^{zr}}{\partial r} + \frac{\partial \sigma^{zz}}{\partial z} + \frac{\sigma^{zr}}{r}
	\end{pmatrix},
\end{equation}
where $\sigma^{\alpha\beta}$ are the components of the stress tensor. Using the notations introduced above, the equation (\ref{eq:momentum-strength-cyl}) can be rewritten as:
\begin{equation}\label{eq:momentum-strength-lite}
	\frac{d\mathbf{U}}{dt} = \frac{1}{\rho}\boldsymbol{\nabla}\cdot\tilde{\boldsymbol{\sigma}} + \frac{1}{r\rho}\begin{pmatrix}
		\sigma^{rr} -  \sigma^{\theta\theta}
		\\
		\sigma^{zr}
	\end{pmatrix},
\end{equation}
where
$$
\tilde{\boldsymbol{\sigma}} = 
\begin{pmatrix} 
	\sigma^{rr} && \sigma^{rz} \\
	\sigma^{zr} && \sigma^{zz}
\end{pmatrix}, \,\,\,\,\boldsymbol{\nabla}\cdot\tilde{\boldsymbol{\sigma}} = \begin{pmatrix}
	\frac{\partial \sigma^{rr}}{\partial r} + \frac{\partial \sigma^{zr}}{\partial z}
	\\
	\frac{\partial \sigma^{zr}}{\partial r} + \frac{\partial \sigma^{zz}}{\partial z}
\end{pmatrix}.
$$
To construct the SPH approximation of the equation of motion, the following identity is used:
$$\boldsymbol{\nabla}\cdot\tilde{\boldsymbol{\sigma}} = \frac{1}{f(r;A)}\left(\boldsymbol{\nabla}\cdot\left(f(r;A)\tilde{\boldsymbol{\sigma}}\right) + \tilde{\boldsymbol{\sigma}}\cdot\boldsymbol{\nabla} f(r;A)\right) - \frac{2f'(r;A)}{f(r;A)}\tilde{\boldsymbol{\sigma}}\cdot\mathbf{e}^r,$$
where $f(r;A)$ is a smooth function of the radial coordinate, depending on $A$ as a parameter.
Using (\ref{eq:scalar-sph-approximation-grad-axisymmetric}), (\ref{eq:vector-sph-approximation-div-axisymmetric}) one can obtain the equation of motion in the form:
\begin{equation}\label{eq:axisymmetrical-general-momentum-strength}
	\frac{d\mathbf{U}_a}{dt} = \sum\limits_{b}\frac{m_b}{2\pi \rho_b\rho_ar_b}\frac{f(r_b;A)}{f(r_a;A)}(\tilde{\boldsymbol{\sigma}}_a + \tilde{\boldsymbol{\sigma}}_b)\cdot\boldsymbol{\nabla}_aW_{ab} - 
	\\
	\frac{2f'(r_a;A)}{\rho_a f(r_a;A)}
	\begin{pmatrix}
		\sigma_a^{rr} \\
		\sigma_a^{zr}
	\end{pmatrix} +
	\frac{1}{r_a\rho_a}\begin{pmatrix}
		\sigma_a^{rr} - \sigma_a^{\theta\theta} \\
		\sigma_a^{zr}
	\end{pmatrix}.
\end{equation}

The equation for internal energy has the following form:
$$\frac{de}{dt} = \frac{1}{\rho}\tilde{\boldsymbol{\sigma}}:(\boldsymbol{\nabla}\otimes\mathbf{U}) + \sigma^{\theta\theta}\frac{U^r}{\rho r}.$$
Let us apply the following identity to approximate $\boldsymbol{\nabla}\otimes\mathbf{U}$:
$$(\boldsymbol{\nabla}\otimes\mathbf{U}) = \frac{\boldsymbol{\nabla}\otimes(f(r;A)\mathbf{U}) - \boldsymbol{\nabla} f(r;A)\otimes\mathbf{U}}{f(r;A)}$$
and write the internal energy equation in the following form:
\begin{multline}\label{eq:axisymmetrical-general-internal-energy-strength}
	\frac{de_a}{dt} = \sum\limits_b\frac{m_b}{2\pi \rho_b\rho_a r_b}
	\frac{f(r_b;A)}{f(r_a;A)} \frac{\tilde{\boldsymbol{\sigma}}_a + \tilde{\boldsymbol{\sigma}}_b}{2}:\left(\boldsymbol{\nabla}_aW_{ab}\otimes\left(\mathbf{U}_b - \mathbf{U}_a\right)\right) + \frac{\sigma^{\theta\theta}_aU^r_a}{r_a\rho_a}=
	\\
	\sum\limits_b\frac{m_b}{2\pi \rho_b\rho_a r_b}
	\frac{f(r_b;A)}{f(r_a;A)} \left(\mathbf{U}_b - \mathbf{U}_a\right)\cdot\left(\frac{\tilde{\boldsymbol{\sigma}}_a + \tilde{\boldsymbol{\sigma}}_b}{2}\cdot\boldsymbol{\nabla}_aW_{ab}\right) + \frac{\sigma^{\theta\theta}_aU^r_a}{r_a\rho_a}	.
\end{multline}

By combining the equation of motion (\ref{eq:axisymmetrical-general-momentum-strength}) and the equation for internal energy (\ref{eq:axisymmetrical-general-internal-energy-strength}) one can obtain the equation for the total energy:
\begin{multline}\label{eq:axisymmetrical-general-energy-strength}
	\frac{dE_a}{dt} = \sum\limits_b\frac{m_b}{2\pi \rho_b\rho_a r_b}
	\frac{f(r_b;A)}{f(r_a;A)} \left(\mathbf{U}_b + \mathbf{U}_a\right)\cdot\left(\frac{\tilde{\boldsymbol{\sigma}}_a + \tilde{\boldsymbol{\sigma}}_b}{2}\cdot\boldsymbol{\nabla}_aW_{ab}\right) + 
	\\
	\frac{1}{\rho_a}(U^r_a\sigma^{rr}_a+U^z_a\sigma^{zr}_a)\left(-\frac{2f'(r_a;A)}{f(r_a;A)}+\frac{1}{r_a}\right).
\end{multline}

\begin{figure}
	\centering
	\includegraphics[width=0.5\linewidth]{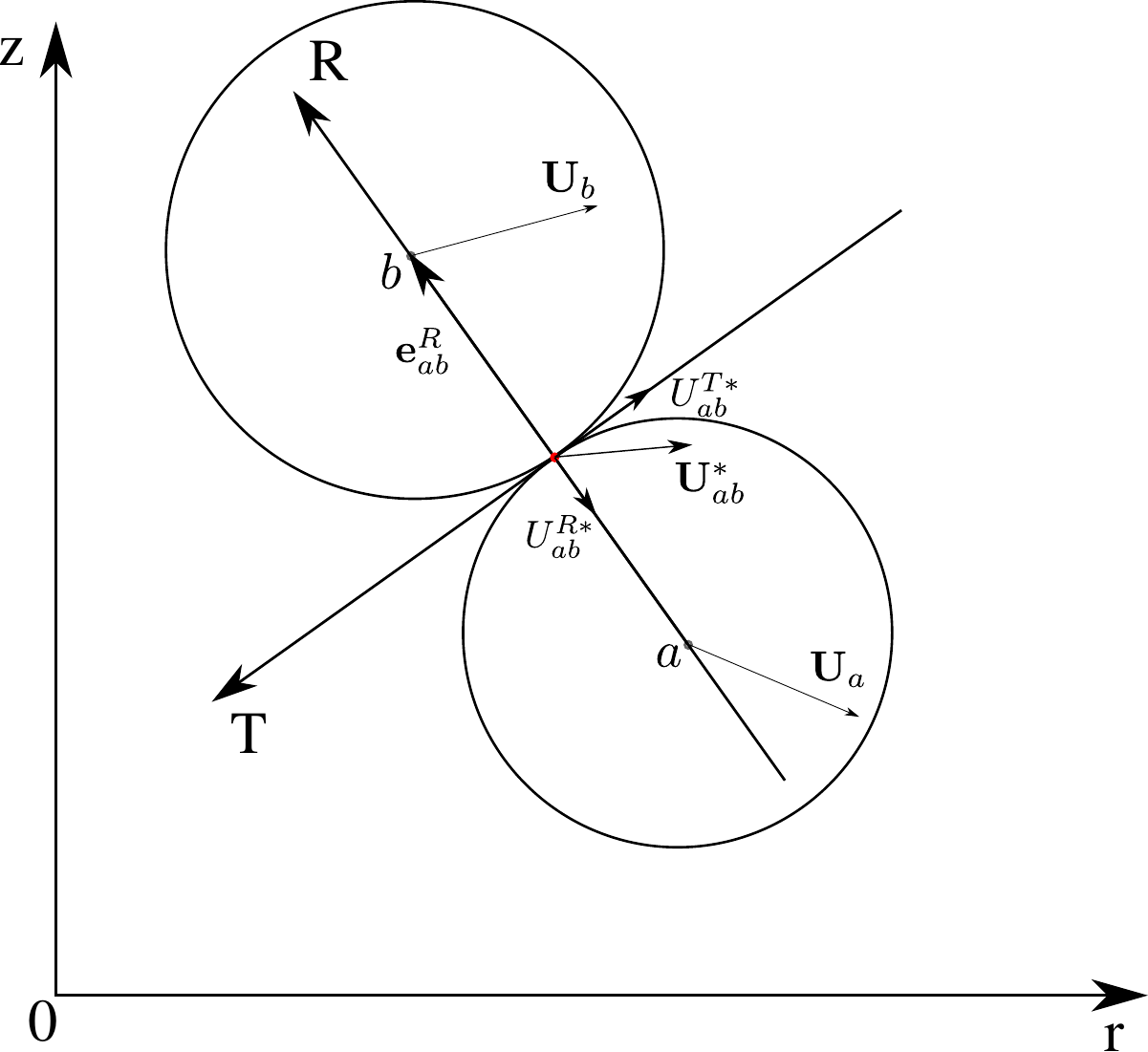}
	\caption{\label{local-coordinate-system} Interaction of particles in the local coordinate system}
\end{figure}
The contact values of velocity and stress are calculated in the local RT coordinate system as shown in Fig.~\ref{local-coordinate-system}.
The general form of the equations of motion and total energy is obtained from Eqs.~(\ref{eq:axisymmetrical-general-momentum-strength}) and (\ref{eq:axisymmetrical-general-energy-strength}) by the substitution described in the~\cite{Parshikov:JCP:2002}:
$$\mathbf{M}^{rz\rightarrow RT}_{ab}\cdot\frac{\mathbf{U}_a + \mathbf{U}_b}{2} \approx \mathbf{U}^{*}_{ab}, \,\,\,\, \mathbf{M}^{rz\rightarrow RT}_{ab}\cdot\left(\frac{\tilde{\boldsymbol{\sigma}}_a + \tilde{\boldsymbol{\sigma}}_b}{2}\cdot\mathbf{e}^R_{ba}\right) \approx \tilde{\boldsymbol{\sigma}}_{ab}^{R*},$$
where $\mathbf{M}^{rz\rightarrow RT}_{ab}$ is the transition matrix from the coordinate system $rz$ to the local coordinate system $RT$:
$$\mathbf{M}^{rz\rightarrow RT}_{ab} = \begin{pmatrix} (e^R_{ba})^r & (e^R_{ba})^z \\
-(e^R_{ba})^z & (e^R_{ba})^r 
\end{pmatrix}, ~~~~ \mathbf{M}^{RT\rightarrow rz}_{ab} = \left(\mathbf{M}^{rz\rightarrow RT}_{ab}\right)^T.$$
The components of the contact velocity and stress vectors are determined in the acoustic approximation by the formulas:
$$U^{*R}_{ab} = \frac{U^R_b\rho_bC_b^l + U^R_b\rho_bC_b^l + \tilde{\sigma}^{RR}_b - \tilde{\sigma}^{RR}_a}{\rho_bC_b^l + \rho_aC_a^l},$$
$$U^{*T}_{ab} = \frac{U^T_b\rho_bC_b^t + U^T_b\rho_bC_b^t + \tilde{\sigma}^{RT}_b - \tilde{\sigma}^{RT}_a}{\rho_bC_b^t + \rho_aC_a^t},$$
$$\tilde{\sigma}^{*RR}_{ab} = \frac{\tilde{\sigma}^{RR}_b\rho_aC_a^l + \tilde{\sigma}^{RR}_a\rho_bC_b^l +  \rho_aC_a^l\rho_bC_b^l(U^R_b - U^R_a)}{\rho_bC_b^l + \rho_aC_a^l},$$
$$\tilde{\sigma}^{*RT}_{ab} = \frac{\tilde{\sigma}^{RT}_b\rho_aC_a^t + \tilde{\sigma}^{RT}_a\rho_bC_b^t +  \rho_aC_a^t\rho_bC_b^t(U^T_b - U^T_a)}{\rho_bC_b^t + \rho_aC_a^t},$$
$$\tilde{\boldsymbol{\sigma}}_{a}^{R} = \mathbf{M}^{rz\rightarrow RT}_{ab}\cdot\left(\tilde{\boldsymbol{\sigma}}_a\cdot\mathbf{e}_{ab}^R\right),$$
$C^l$ is the longitudinal sound speed, $C^t$ is the transversal sound speed.

As a result:
\begin{equation}\label{eq:axisymmetrical-general-momentum-strength-Rieman}
	\frac{d\mathbf{U}_a}{dt} = -\sum\limits_{b}\frac{m_b}{2\pi \rho_b\rho_ar_bh_{ab}}\frac{f(r_b;A)}{f(r_a;A)}2\mathbf{M}^{RT\rightarrow rz}_{ab}\cdot\tilde{\boldsymbol{\sigma}}_{ab}^{R*}W'_{ab} - 
	\\
	\frac{2f'(r_a;A)}{\rho_a f(r_a;A)}
	\begin{pmatrix}
		\sigma_a^{rr} \\
		\sigma_a^{zr}
	\end{pmatrix} +
	\frac{1}{r_a\rho_a}\begin{pmatrix}
		\sigma_a^{rr} - \sigma_a^{\theta\theta} \\
		\sigma_a^{zr}
	\end{pmatrix},
\end{equation}

\begin{equation}\label{eq:axisymmetrical-general-energy-strength-Rieman}
	\frac{dE_a}{dt} = -\sum\limits_b\frac{m_b}{2\pi \rho_b\rho_a r_bh_{ab}}
	\frac{f(r_b;A)}{f(r_a;A)} 2\mathbf{U}^{*}_{ab}\cdot\tilde{\boldsymbol{\sigma}}_{ab}^{R*} W'_{ab} + 
	\\
	\frac{1}{\rho_a}(U^r_a\sigma^{rr}_a+U^z_a\sigma^{zr}_a)\left(-\frac{2f'(r_a;A)}{f(r_a;A)}+\frac{1}{r_a}\right).
\end{equation}
To ensure conservativity, the conditions 1 and 2 on the function $f(r,A)$ are imposed, assuming that in Eqs. (\ref{eq:axisymmetrical-general-momentum-strength-Rieman}), (\ref{eq:axisymmetrical-general-energy-strength-Rieman}) $A = r_a$ and $f(r_a,r_a) = 1$. The function $f(r;r_a)$ is chosen again in the form (\ref{eq:g-function}). Then the ``harmonic mean'' scheme (the function $g$ for this scheme is given in the table \ref{table:axisym-schemes-examples}) is applied:

\begin{equation}\label{eq:axisymmetrical-general-momentum-strength-Rieman-harmonic}
	\frac{d\mathbf{U}_a}{dt} = -\sum\limits_{b}\frac{m_b}{2\pi \rho_b\rho_ah_{ab}}\frac{r_b+r_a}{r_ar_b}\mathbf{M}^{RT\rightarrow rz}_{ab}\cdot\tilde{\boldsymbol{\sigma}}_{ab}^{R*}W'_{ab}  -
	\frac{\sigma_a^{\theta\theta}}{r_a\rho_a}\mathbf{e}^r,
\end{equation}

\begin{equation}\label{eq:axisymmetrical-general-energy-strength-Rieman-harmonic}
	\frac{dE_a}{dt} = -\sum\limits_b\frac{m_b}{2\pi \rho_b\rho_a h_{ab}}
	\frac{r_b+r_a}{r_ar_b}\mathbf{U}^{*}_{ab}\cdot\tilde{\boldsymbol{\sigma}}_{ab}^{R*} W'_{ab}.
\end{equation}

The elastic stress tensor-deviator is defined as:
\begin{equation}\label{stress-tensor-evolution}
	\begin{gathered}
		\dot{S}_{rr} = 2G\dot{e}_{rr} - 2\omega S_{rz},
		\\
		\dot{S}_{zz} = 2G\dot{e}_{zz} + 2\omega S_{rz},
		\\
		\dot{S}_{\theta\theta} = 2G\dot{e}_{\theta\theta},
		\\
		\dot{S}_{rz} = 2G\dot{e}_{rz} + \omega (S_{rr} - S_{zz}),
	\end{gathered}
\end{equation}
where $G$ is the shear modulus, $\omega$ is the angular velocity:
\begin{equation}\label{rotation-rate}
	\omega = \frac{1}{2}\left( \frac{\partial U^z}{\partial r} + \frac{\partial U^r}{\partial z}\right).
\end{equation}
The strain rate tensor-deviator is defined as:
\begin{equation}\label{stress-tensor-evolution}
	\begin{gathered}
		\dot{e}_{rr} = \frac{1}{3}\left(2\frac{\partial U^r}{\partial r} - \frac{\partial U^z}{\partial z} - \frac{U^r}{r}\right),
		\\
		\dot{e}_{zz} = \frac{1}{3}\left(2\frac{\partial U^z}{\partial z} - \frac{\partial U^r}{\partial r} - \frac{U^r}{r}\right),
		\\
		\dot{e}_{\theta\theta} = \frac{1}{3}\left(2\frac{U^r}{r} - \frac{\partial U^r}{\partial r} - \frac{\partial U^z}{\partial z}\right),
		\\
		\dot{e}_{rz} = \frac{1}{2}\left(\frac{\partial U^z}{\partial r} + \frac{\partial U^r}{\partial z}\right).
	\end{gathered}
\end{equation}

An estimate of the first derivatives of the velocity components is:
\begin{equation}\label{eq:velocity-derivative-sph-cyl}
	\boldsymbol{\nabla}\otimes\mathbf{U}_a = \sum\limits_b\frac{m_b}{\pi r_b\rho_b} \boldsymbol{\nabla}_aW_{ab}\otimes \left(\mathbf{U}^{*}_{ab} - \mathbf{U}_a\right).
\end{equation}
Substituting the estimates of the velocity derivatives from Eq.~(\ref{eq:velocity-derivative-sph-cyl}) into Eqs.~(\ref{stress-tensor-evolution}), (\ref{rotation-rate}), one can  obtain expressions for the strain rate tensor-deviator and angular velocity.

To perform simulations  the first-order explicit Euler scheme for time integration is used.

The schemes presented in Table \ref{table:axisym-schemes-examples} are fully conservative, i.e., they preserve the total energy and the balance between the kinetic and internal energies of the medium, which is important for solving problems where numerical heating is not allowed.

\subsection{Demonstrating the conservativity of the proposed scheme}

To demonstrate the conservativity of the proposed scheme, the modeling of the cylindrical Verni problem is performed with the original A.~N.~Parshikov~\cite{Parshikov2000} scheme and with the ``harmonic mean'' scheme. In this problem, the radial motion of a cylindrical shell toward the axis of symmetry is modeled until it comes to a full stop. The modeling is carried out using two-dimensional axisymmetric formulation. The periodic boundary condition (infinitely long cylindrical shell) is set along the $z$-axis.

The shell material is modeled using the equation of state in the form:
$$P = c_0^2(\rho - \rho_0) + (\gamma-1) \rho e.$$ 

The initial outer and inner radii of the shell are set to $R_1 = 0.1\,$m and $R_0 = 0.08\,$m as in Ref.~\cite{Howell2002}. The shell material is aluminum with the density $\rho = 2785\,$kg/m$^3$, the shear modulus $G = 27.6\,$GPa, the yield strength $Y = 0.3\,$GPa, the bulk sound speed $c_0 = 5328\,$m/s, Gr\"uneisen parameter $\gamma = 2$.

\begin{figure}[t]
	\includegraphics[width=\linewidth]{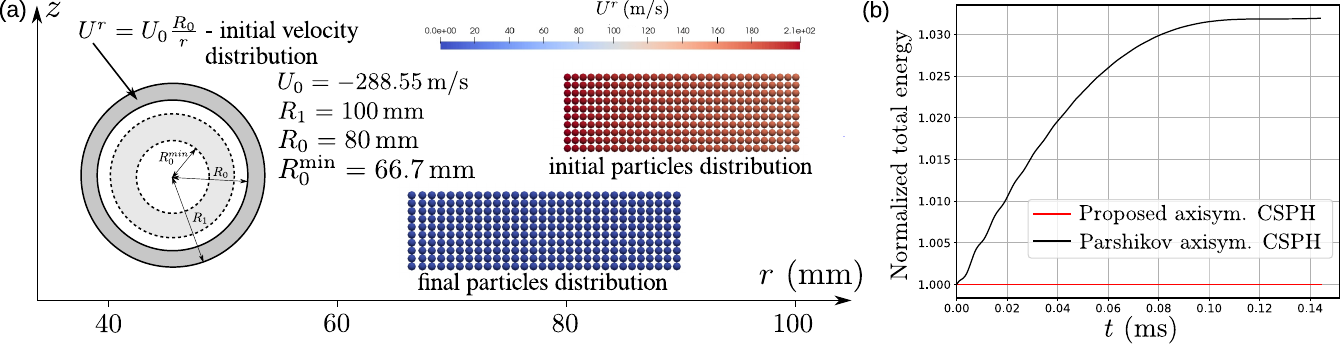}
	\caption{\label{verni} (a) Problem setup and SPH particles distribution in the cylindrical Verni problem. (b) Evolution of the total energy in the cylindrical Verney problem using the proposed axisymmetric CSPH ``harmonic mean'' scheme and the A.N. Parshikov~\cite{Parshikov2000} axisymmetric CSPH scheme.}
\end{figure}  

Initial velocity distribution of SPH particles is
\begin{equation}\label{eq:VernyU}
	U^r(r) = U_0\frac{R_0}{r},
\end{equation}
where $U_0 = 208.55\,$m/s.

The velocity $U_0$ required to reach the inner radius at the first stop of the inner boundary of the cylindrical shell $r_0$ is determined by the formula (see, e.g., \cite{Howell2002}):
$$U_0 = \sqrt{\frac{2YF(\alpha,\lambda)}{\sqrt{3}\rho\ln(R_1/R_0)}},$$
where
$$F(\alpha, \lambda) = \int\limits_{\lambda}^1x\ln\left(1+\frac{2\alpha + \alpha^2}{x^2}\right)dx,$$
$$\alpha = \frac{R_1-R_0}{R_0},\,\,\lambda=\frac{r_0}{R_0}.$$
One may obtain that for the given parameters the inner radius corresponding the full stop is $r_0 = 0.0667\,$m.

The inner radius at the moment of the full stop of the shell proposed by axisymmetric CSPH scheme with ``harmonic mean'' is $0.06678\,$m, while A.~N.~Parshikov~\cite{Parshikov2000} axisymmetric CSPH scheme provides the value of $0.06638\,$m.
Fig.~\ref{verni}(a) shows the initial packing of SPH particles and the radial velocity is distributed according to the law~(\ref{eq:VernyU}) and the final distribution of SPH particles.

Fig.~\ref{verni}(b) shows the total energy change in the cylindrical Verni problem with the axisymmetric CSPH schemes: original A.~N.~Parshikov~\cite{Parshikov2000} and the proposed ``harmonic mean'' (current work) ones. The proposed scheme exactly preserves the total energy, while the A.N. Parshikov~\cite{Parshikov2000} scheme demonstrate the total energy growth by more than three percent, i.e. numerical heating takes place.

In order to speed up calculations and to eliminate local unphysical features of the flow, it is reasonable to use the procedures presented further.

\section{Possible practical scheme improvements violating the conservativity}
\label{BC_and_particle_adoptation}

\subsection{Boundary condition near the axis of symmetry}

For particles near the axis of symmetry the support radius of the smoothing kernel function is incomplete, since there are no neighboring particles at $r<0$. To solve this problem, it is necessary to introduce a special boundary condition here by providing ghost particles, which are obtained by reflection of original particles across the symmetry axis, to the calculation of SPH sums.

The ``harmonic mean'' scheme allows the reflection of particles across the axis, but numerical calculations show that the best result is achieved with the transition to the A.N. Parshikov~\cite{Parshikov2000} scheme near the axis of symmetry:
$$\frac{d\varepsilon_a}{dt} = -\sum\limits_{b}\frac{m_b}{2\pi r_b\rho_b h_{ab}}2(U_{ab}^{R*} - U_a^R)W'_{ab} + \frac{U^r_a}{r_a},$$ 
$$\frac{d\mathbf{U}_a}{dt} = -\sum\limits_{b}\frac{m_b}{2\pi r_b \rho_b\rho_ah_{ab}}2\mathbf{M}^{RT\rightarrow rz}_{ab}\cdot\tilde{\boldsymbol{\sigma}}_{ab}^{R*}W'_{ab}  +
\frac{\sigma_a^{rr} - \sigma_a^{\theta\theta}}{r_a\rho_a}\mathbf{e}^r +\frac{\sigma_a^{rz}}{r_a\rho_a}\mathbf{e}^z, $$
$$\frac{dE_a}{dt} = -\sum\limits_b\frac{m_b}{2\pi r_b\rho_b\rho_a h_{ab}}
2\mathbf{U}^{*}_{ab}\cdot\tilde{\boldsymbol{\sigma}}_{ab}^{R*} W'_{ab} + \frac{\sigma_a^{rr}U^r_a + \sigma_a^{rz}U^z_a}{r_a\rho_a}.$$
Thus, the interaction of the base particle with the ghost particles reflected across the symmetry axis is allowed, but the conservativity of the total energy and total momentum is violated.

The switch to the A.~N.~Parshikov scheme~\cite{Parshikov2000} is made only for particles closer than twice smoothing length $h$ to the axis $z$ ($r < 2h$). Interactions with the ghost particles are added to the SPH sums, and their velocities, coordinates, and deviatoric stress tensor components should be properly transformed.

\subsection{Particle splitting}
The key target properties of the particle splitting algorithm are:
1) conservativity (total momentum, total energy and total angular momentum must be conserved before and after splitting), 2) new particles should not be located close to their neighboring particles (particles should fill the computational domain as uniformly as possible).

Let $D_{split}$ be the target value of particle size at splitting. The splitting condition may be introduced as: $|D_a/2 - D_{split}| < |D_a - D_{split}|$.

The following algorithm is used to split particles:
\begin{enumerate}
	\item When the splitting condition is met, the particle is split into $4$ particles. The particles are located at the vertices of a square with side $D/2$, where $D$ is the size of the particle before splitting (see Fig.~\ref{refining-pattern}(a)). The geometric center of the square is located at the point $\vec{r}_a$, where the mother particle with number $a$ is located.
	\item The square is rotated so that the segment connecting the centers of the mother particle and its nearest neighbor is perpendicular to one of the sides of the square.
	\item After splitting, the particles obtain the size $D' = D_a/2$, the mass $m' = m_a/4$, and the density $\rho' = m'/(2\pi r\left.D'\right.^2)$, while the other values (the velocity, the specific internal energy, etc.) remain unchanged.
\end{enumerate}
Interpolation of physical quantities to the positions of daughter particles is not performed to secure conservativity. It should be noted that the described procedure of particle splitting preserves conservativity of the schemes considered above.

\subsection{Particle merging}
 
\begin{figure}
	\centering
	\includegraphics[width=0.6\linewidth]{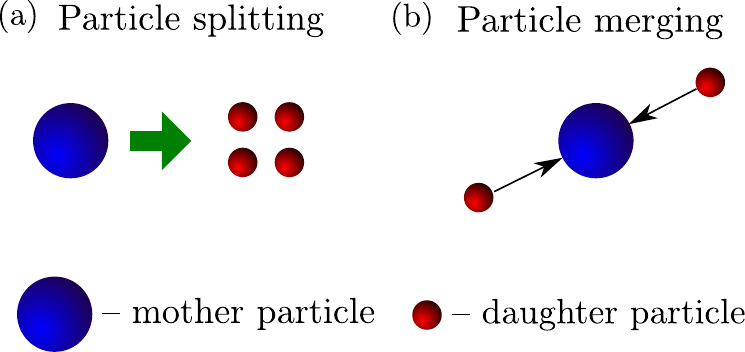}
	\caption{\label{refining-pattern} Patterns for splitting and merging of SPH particles}
\end{figure}

Particle merging is performed by pairs of SPH particles as shown in Fig.~\ref{refining-pattern}(b). When merging particles it is necessary to follow the same principles as for splitting. However, it is no longer possible to satisfy exactly the law of energy conservation without artificial heating of the particles, since in the general case the merging of two particles would be similar to a completely inelastic impact:
$$m_a\frac{(U_a^r)^2 + (U_a^z)^2}{2} + m_b\frac{(U_b^r)^2+(U_b^z)^2}{2} \neq (m_a+m_b)\frac{({U'}^r)^2 + ({U'}^z)^2}{2},$$
where $\mathbf{U'} = (m_a\mathbf{U}_a + m_b\mathbf{U}_b)/(m_a + m_b)$ is the velocity of the coupled SPH particle.
The kinetic energy defect (see~\cite{Sommerfeld1944}) is equal to
$$m_a\frac{(U_a^r)^2 + (U_a^z)^2}{2} + m_b\frac{(U_b^r)^2+(U_b^z)^2}{2} - (m_a+m_b)\frac{({U'}^r)^2 + ({U'}^z)^2}{2} = - \frac{m_am_b}{2(m_a + m_b)}\left(\mathbf{U}_a - \mathbf{U}_b\right)^2 < 0.$$
This defect of kinetic energy must either be converted into internal energy or excluded from the calculation. To avoid artificial heating here the second option is chosen. But if the velocities of particles are equal the defect of kinetic energy is zero, and the total kinetic energy is conserved.

Let $D_{merge}$ be the target value of the merged particle size. Merging of particles $a$ and $b$ is performed under the condition:
$$\left|\sqrt{D_a^2 + D_b^2}-D_{merge}\right| < \left|D_a - D_{merge}\right|,$$ 
$$\left|\sqrt{D_a^2 + D_b^2}-D_{merge}\right| < \left|D_b - D_{merge}\right|.$$

After merging, the size of the new particle is $D' = \sqrt{D_a^2 + D_b^2}$ (the values related to the merged particle are indicated by a dash). The mass of the new particle is $m' = m_a + m_b$. The coordinate is $\mathbf{r'} = (m_a\mathbf{r}_a + m_b\mathbf{r}_b)/(m_a + m_b)$. The density is $\rho' = m'/(2\pi r' D'^2)$.  Velocity is $\mathbf{U'} = (m_a\mathbf{U}_a + m_b\mathbf{U}_b)/(m_a + m_b)$. The specific internal energy is $e' = (m_ae_a + m_be_b)/(m_a + m_b)$.

The described procedures violate the conservativity of the scheme a bit, as shown by the test simulation of the collision of two liquid tin drops.

\subsection{Collision of two drops of liquid tin}
The proposed scheme of the axisymmetric CSPH method is conservative, but the boundary condition near the symmetry axis described above violates the conservativity, which leads to a change in the total momentum and total energy. To estimate the change of total energy and total momentum associated with the boundary condition near the symmetry axis, consider the problem of collision of two liquid tin droplets of different sizes. The initial radii of the droplets are $0.1\,$m and $0.05\,$m. The initial velocities of the droplets are $U_0 = 200\,$m/s and oppositely directed along the $z$-axis. The initial sizes of SPH-particles are equal to $2\,$mm.

\begin{figure}
	\centering
	\includegraphics[width=\linewidth]{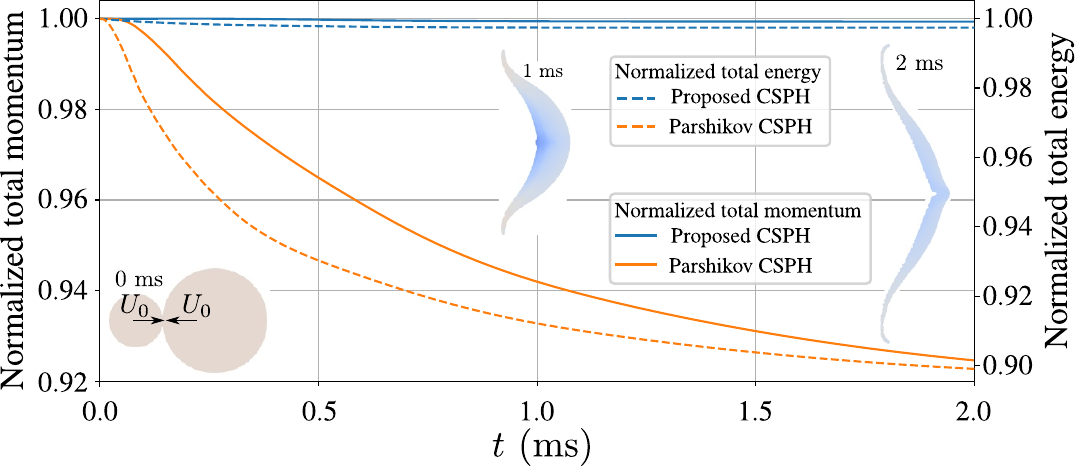}
	\caption{\label{fig:conservation}Evolution of total momentum and total energy in the collision problem of two liquid tin droplets using the proposed axisymmetric CSPH ``harmonic mean'' scheme and the scheme of the axisymmetric CSPH method of A.N. Parshikov~\cite{Parshikov2000}.}
\end{figure}

Fig.~\ref{fig:conservation} shows the evolution of the total momentum and total energy when using the proposed axisymmetric CSPH ``harmonic mean'' scheme and the scheme of the axisymmetric CSPH method of A.N. Parshikov \cite{Parshikov2000}. The presented data show that the scheme proposed in the present work preserves the total momentum and total energy much more accurately, since in general it is conservative, and the introduced errors are of local character near the symmetry axis.

\section{Numerical examples}
\label{Numerical-examples}
\subsection{Axisymmetric Sod's test}
Here the cylindrical axisymmetric Riemann problem in an ideal gas is considered. The simulation is carried out in the 2D axisymmetric formulation with the smoothing kernel Wendland~C$^2$. The smoothing length parameter is $\theta = 0.5$. The initial particle size is $D_0 = 1.41\,$mm. The initial conditions are given as follows:
\begin{equation*}
	\rho = 
	\begin{cases}
		1, \, r \leq 0.4,
		\\
		0.125, \, r > 0.4,
	\end{cases} 
\end{equation*} 
\begin{equation*}
	p = 
	\begin{cases}
		1, \, r \leq 0.4,
		\\
		0.1, \, r > 0.4.
	\end{cases} 
\end{equation*} 
At the initial moment of time the gas is at rest. The evolution of gas is simulated until $0.25\,$ms. The ideal gas equation of state is used:
\begin{equation}\label{ideal-gas}
	P = (\gamma-1)\rho e,
\end{equation}
with $\gamma = 1.4$.
\begin{figure}
	\centering
	\includegraphics[width=\linewidth]{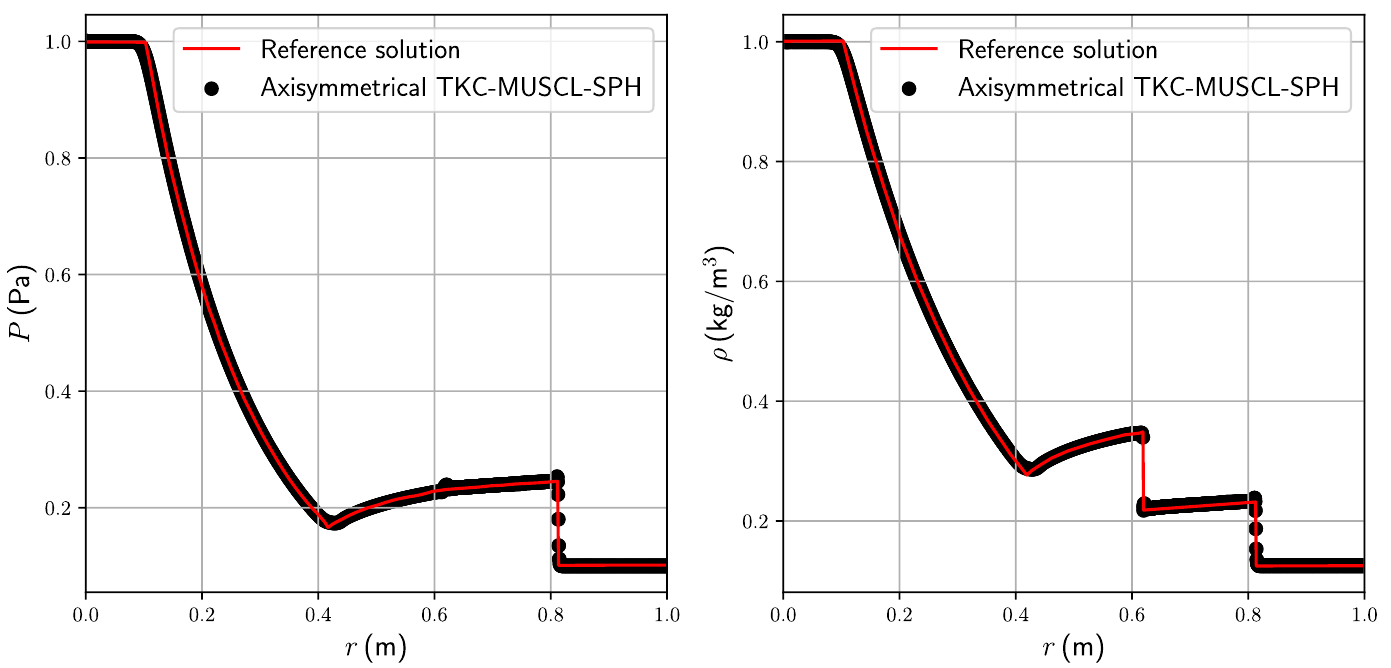}
	\caption{\label{SOD} The axisymmetric Sod's test: the gas, initially compressed at $r \leq 0.4$, begin to expand with shock wave to the outer region, while the rarefaction wave propagates to the center $r = 0$. }
\end{figure}

Fig.~\ref{SOD} shows the density and pressure profiles obtained by the proposed axisymmetric CSPH method with the TKC kernel gradient correction procedure (see Appendix) and piecewise linear reconstruction of the values at the contact using the MUSCL scheme (TKC-MUSCL-SPH). The reference solution is taken from Ref.~\cite{Toro2013}.

\subsection{Sedov's explosion test}

\begin{figure}
	\centering
	\includegraphics[width=\linewidth]{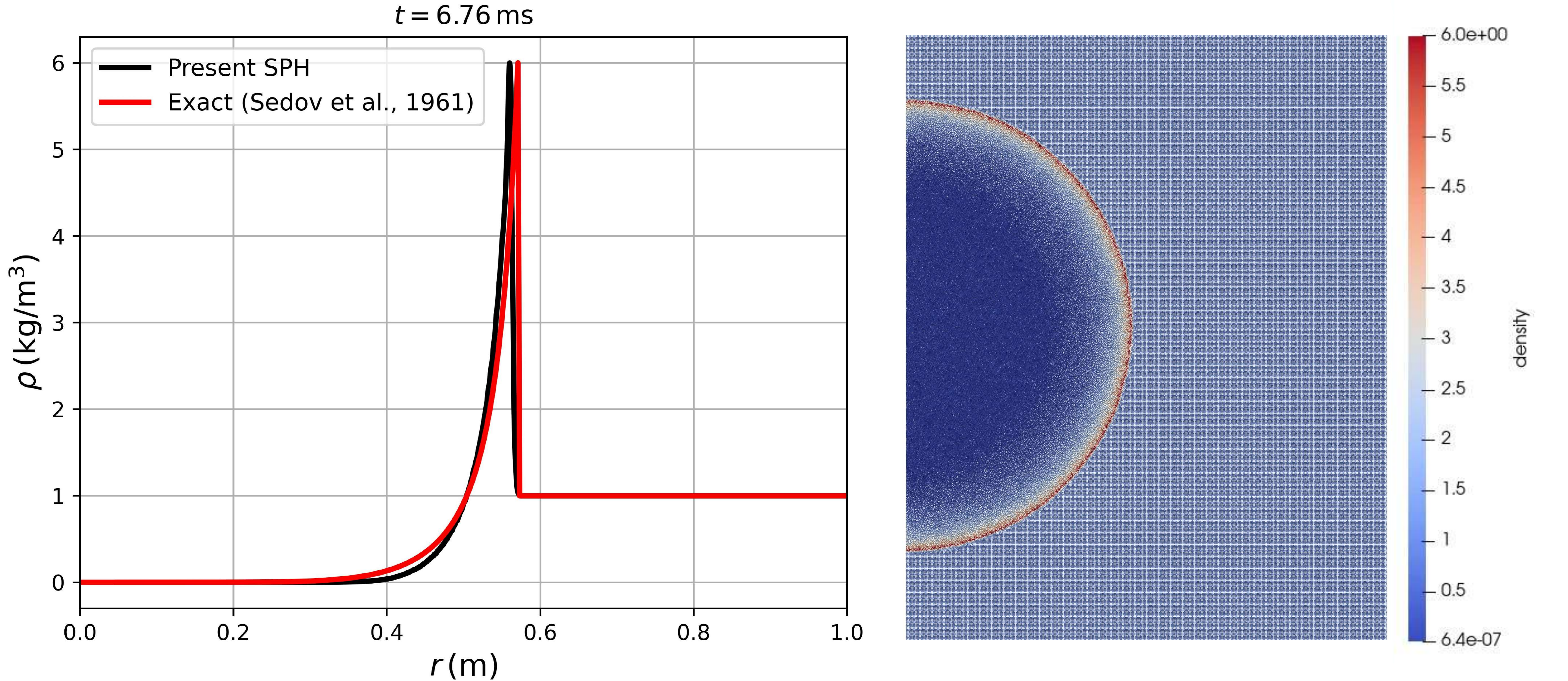}
	\caption{\label{Sedov}Sedov point explosion test. Simulation results are shown for the moment $t=6.76
	\,$ms.}
\end{figure}

This problem demonstrates the accuracy of modeling a spherically diverging from the center shock wave in the axisymmetric case. The ideal gas with an equation of state (\ref{ideal-gas}) and $\gamma=1.4$ is set the 2D axisymmetric region $0<r<1\,$m, $-1\,\text{m} < z < 1\,$m. The initial density of the gas is $\rho_0 = 1\,$kg/m$^3$. In the region $\sqrt{r^2+z^2}\leq 0.03\,$m at the initial time moment, the internal energy of the gas is $e_1 = 10^7\,$J/kg. The internal energy of the remaining gas is $e_0 = 0.1\,$J/kg. The gas is initially at rest. The formulation and analytical solution for the problem was provided in Ref.~\cite{Sedov1959}. 

The maximum reached compression at the shock front is 6. The dependence of the shock wave position $R$ on time is given by the formula:
\begin{equation}\label{sedov-R-t}
	R = 1.033\left(\frac{Et^2}{\rho_0}\right)^{1/5},
\end{equation}
where $E = 1131\,$J is the total energy of the explosion.

Fig.~\ref{Sedov} shows the density distribution obtained by the TKC-CSPH method with $\theta = 0.55$ at time $t = 6.76\,$ms. The splitting/merging and SPH particles shifting algorithms Ref.~\cite{Michel2022} are used in the simulations. The Roe approximation with diffusion limiter from Ref.~\cite{Meng2021} is used as contact velocity and pressure values. The presented data show that the symmetry of the spherically divergent wave is preserved, and the shape and position of the shock front agrees with the analytical solution well.

\subsection{Teylor bar test}

The simulations of impact of a metal cylinder with a rigid wall are usually performed to validate the elastic-plastic models and numerical schemes. Here the Jonson and Cook plasticity model~\cite{Johnson1983} is used, where yield stress $Y$ is expressed as:
$$Y = (A+B\epsilon^n)(1+C\ln(\dot{\varepsilon}^*))(1-\left.T^*\right.^m),$$
where $\epsilon$ is the equivalent plastic strain, $\dot{\varepsilon}^* = \dot{\varepsilon}/\dot{\varepsilon}_0$ is the dimensionless plastic strain rate, $\dot{\varepsilon}_0\,s^{-1}$, $T^*$ is the homologous temperature. 

The experimental setup and model parameters are taken from Ref.~\cite{Johnson1983}. Table~\ref{table:Teylor} summarizes the numerical experimental results and compares the obtained final compression of metallic cylinders with the computational results from the paper~\cite{Johnson1983}. Modeling is carried out by the proposed axisymmetric CSPH method with piecewise linear reconstruction of values at the contact (MUSCL-SPH).

\begin{figure}[t]
	\centering
	\includegraphics[width=\linewidth]{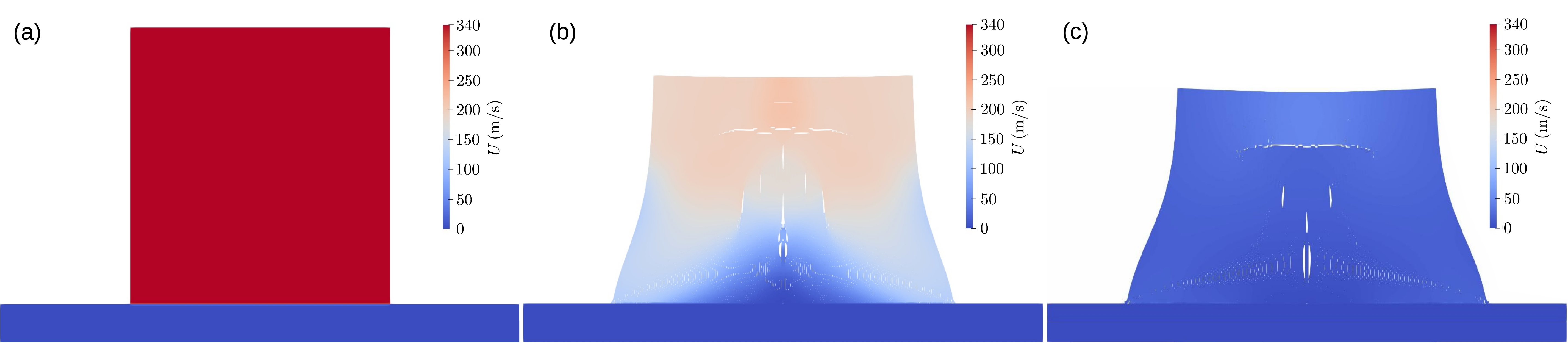}
	\caption{\label{teylor-steel} Modeling the impact of a steel cylinder on a solid wall. The initial length of the cylinder is $L_0 = 0.81\,$cm, the diameter is $D_0 = 0.762\,$cm. The velocity of impact is $343\,$m/s. (a) Start of impact $t=0$, (b) $t=5\,\mu$s, (c) $t=11\,\mu$s.}
\end{figure}

\begin{figure}[t]
	\centering
	\includegraphics[width=\linewidth]{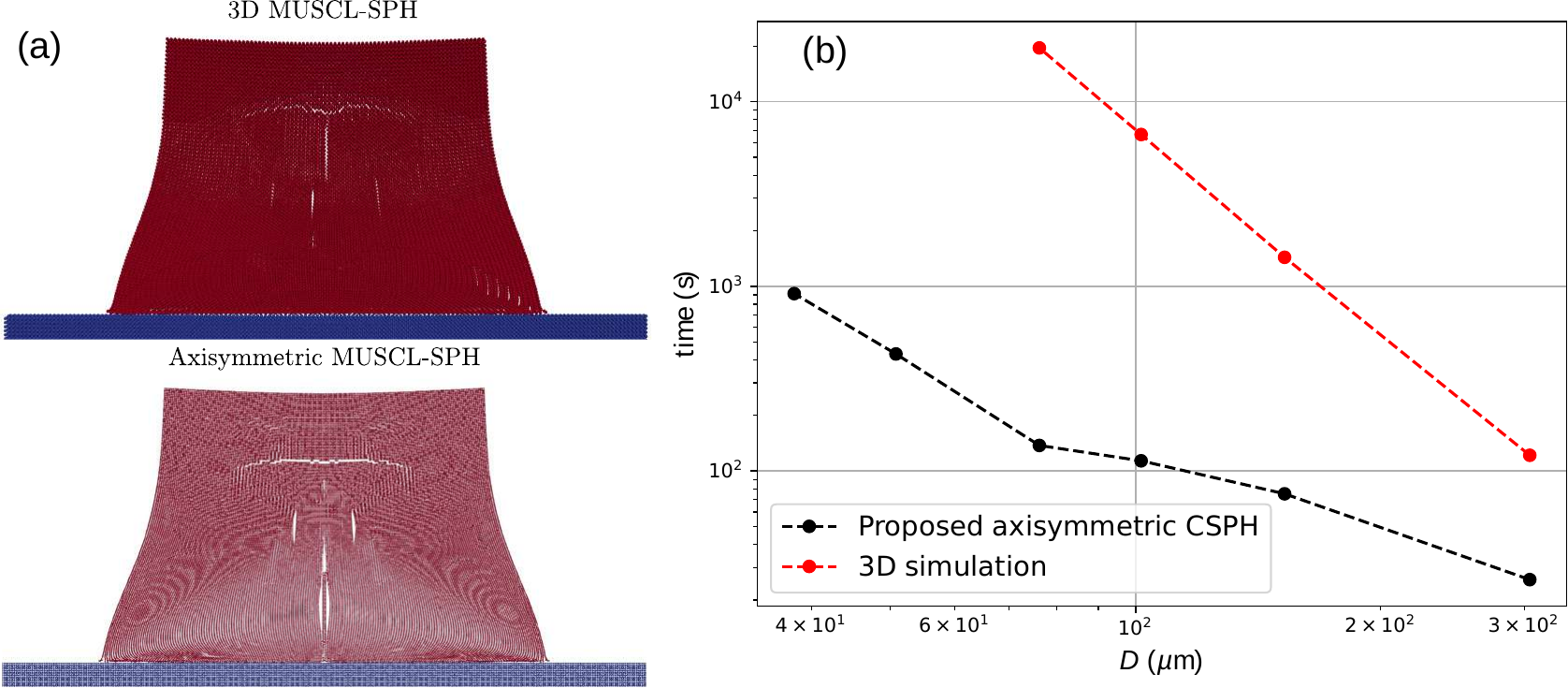}
	\caption{\label{2D-3D-simulation-time} (a) Comparison of the modeling results of a steel cylinder impact on a solid wall using the three-dimensional MUSCL-SPH method in Cartesian coordinates and the proposed axisymmetric MUSCL-SPH method. The initial length of the cylinder is $L_0 = 0.81\,$cm, the diameter is $D_0 = 0.762\,$cm. The velocity of impact is $343\,$m/s.  The moment of time $t = 11\,\mu$s is shown. 
		(b) The comparison of simulation times involves a steel cylinder impacting a rigid wall using the MUSCL-SPH method at different particle sizes in both 2D axisymmetric and 3D Cartesian cases on 16 computational cores. The impact speed is $343\,$m/s, the initial length of the cylinder is $L_0 = 0.81\,$cm, and the calculation ran until $t = 11\,\mu$s.}
\end{figure}

Fig.~\ref{teylor-steel} shows the initial, intermediate, and final phases of the impact of a steel cylinder with a rigid wall. Fig.~\ref{2D-3D-simulation-time} compares the results obtained via the three-dimensional MUSCL-SPH method in Cartesian coordinates and the proposed axisymmetric MUSCL-SPH method at time $t = 11\,\mu$s. In the three-dimensional case there is a central thin slice is shown. The initial particle size in the three-dimensional case is $75.6\,\mu$m, the total number of particles is 1264190, and the simulation time on 128 cores is 1 hour 13 minutes.  The initial particle size in the axisymmetric case is $40.5\,\mu$m, the total number of particles is 24064, and the simulation time on 8 cores is 9 minutes. Fig.~\ref{2D-3D-simulation-time} compares simulation times for the same problem (simulation up to $t = 11\,\mu$s) at different particle sizes in two-dimensional axisymmetric and three-dimensional Cartesian cases, utilizing 16 computational cores.

\begin{table}[H]
	\center{
		\caption{\label{table:Teylor} Initial length and final compressions of the specimens in the Taylor test. The initial diameter of all specimens is $D_0 = 0.762\,$cm.}
		\tabcolsep8pt
		\renewcommand{\arraystretch}{1.25}
		\begin{tabular}{|c|c|c|c|c|}
			\hline
			Material    & $L_0,\,$cm & $U_0,\,$m/s & $L/L_0$          & $L/L_0$  \\
			&            &             & {\small (Jonson \& Cook)} & {\small (Proposed method)} \\
			\hline
			{\small OFHC COPPER} & $2.54$	 & $130$       & $0.811$          & $0.81$		         \\
			\hline
			{\small OFHC COPPER} & $2.54$	 & $146$       & $0.776$          & $0.778$		         \\
			\hline
			{\small OFHC COPPER} & $2.54$	 & $190$       & $0.678$          & $0.69$		         \\
			\hline
			
			{\small ARMCO IRON} & $2.54$	     & $197$       & $0.793$          & $0.81$		         \\
			\hline
			{\small ARMCO IRON} & $2.54$	     & $221$       & $0.754$          & $0.77$		         \\
			\hline
			{\small ARMCO IRON} & $1.26$		 & $279$       & $0.662$          & $0.72$		         \\
			\hline
			
			{\small 4340 STEEL} & $2.54$	     & $208$       & $0.885$          & $0.878$		         \\
			\hline
			{\small 4340 STEEL} & $1.27$		 & $282$       & $0.812$          & $0.82$		         \\
			\hline
			{\small 4340 STEEL} & $0.81$		 & $343$       & $0.748$          & $0.77$		         \\
			\hline
		\end{tabular}
		\tabcolsep15pt
	}
\end{table}

\section{Shock wave weakening by a breakaway sand barrier}
\label{SW-weakening}
Low-strength breakaway barriers made of a mixture of sand and cement can be used to protect against shock waves in the air generated by the detonation of an explosive. The possibility of shock wave weakening by breakaway barriers has been investigated experimentally in Refs.~\cite{Golub2005,Golovastov2024,Mirova2015} and others.

\subsection{The problem of explosion in a sand cylinder}
\begin{figure}[t]
	\centering
	\includegraphics[width=0.8\textwidth]{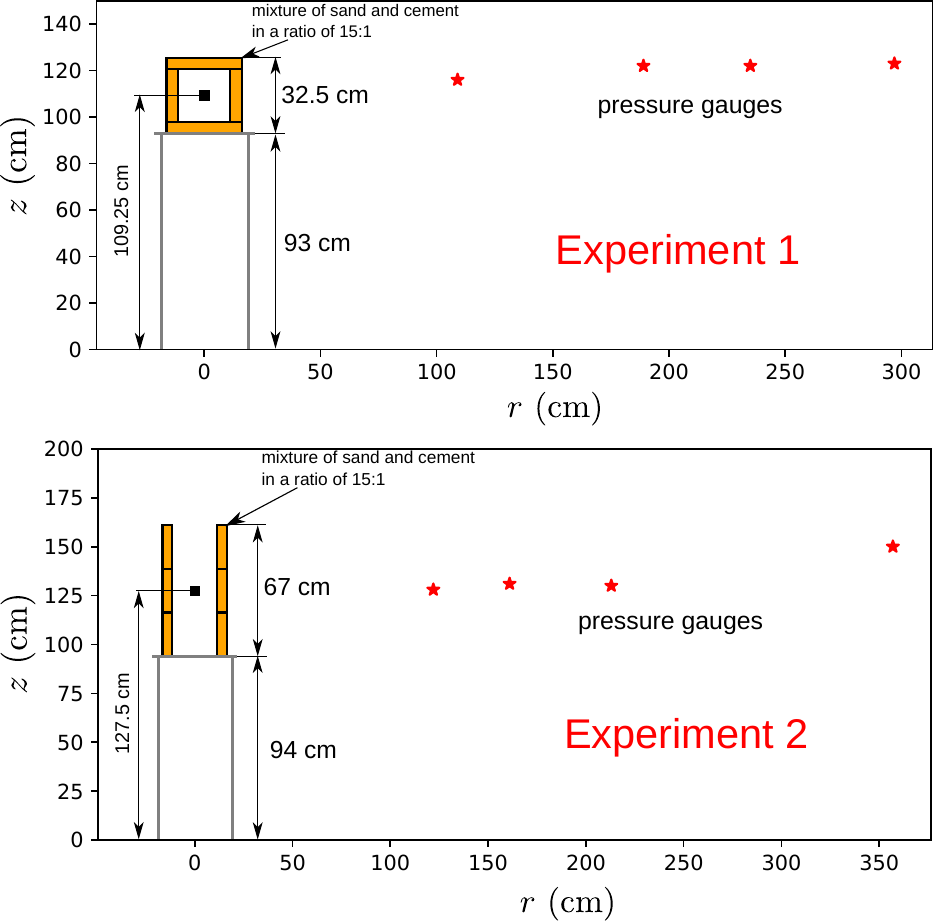}
	\caption{\label{fig:explosion-in-sand-setup} Problem formulations of explosion inside a sand cylinder. Slices in the vertical plane are shown}
\end{figure}

Using the axisymmetric TKC-CSPH method with SPH particle splitting/merging and shifting algorithms, the explosion of the protective sand cylinder by the HE charge is simulated. Four experiments are simulated: two air explosion experiments and two explosion experiments inside the protective sand cylinder~\cite{Golub2005}. Two types of protective sand cylinders are considered. The problem setup for both cases is shown in Fig.~\ref{fig:explosion-in-sand-setup}. In the first case, the cylinder has the inner diameter of $22.5\,$cm and the outer diameter of $32.5\,$cm. The cylinder is closed at the top and bottom by $5\,$cm thick sand covers. In the second case, the cylinder consists of three rings with an outer diameter of $33\,$cm and an inner diameter of $22\,$cm. The material density of the sand cylinder is 1674 kg/m$^3$. In the numerical simulation of these experiments the explosion chamber is represented by a rigid steel cylinder of $3.66\,$m height and $3.7\,$m radius.

\begin{figure}[t]
	\centering
	\includegraphics[width=\textwidth]{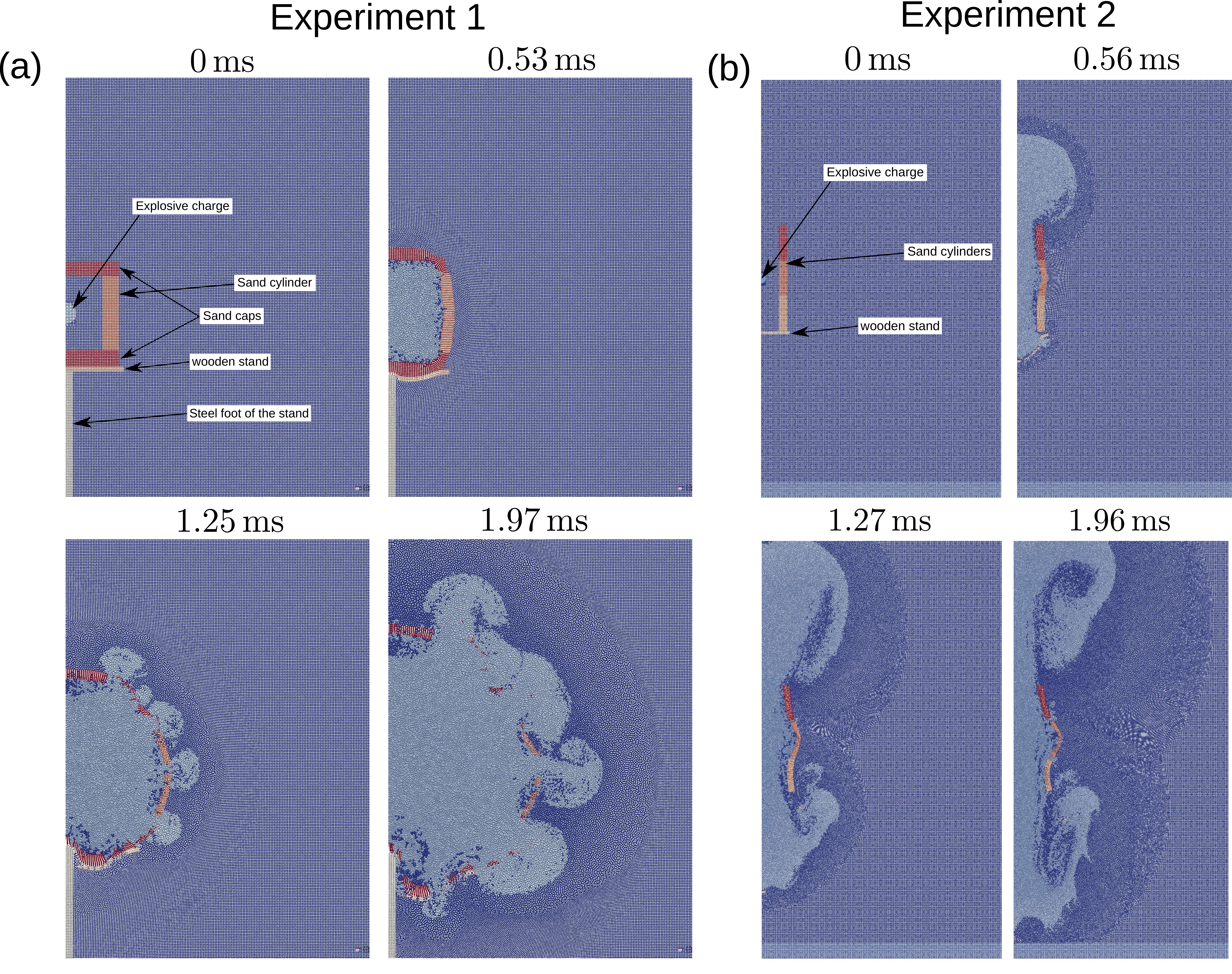}
	\caption{\label{fig:explosion2-in-sand} Explosion inside a sand cylinder: dynamics of sand barrier failure in the (a) first and (b) second experiment. The figures show the distribution of materials at different moments of time.}
\end{figure}

The ideal gas equation of state with $\gamma=1.4$ is used to model air.
The JWL~\cite{Banks2008} equation of state for C-4 is used to model detonation products (constants for the equation of state are taken from Ref.~\cite{Urtiew2008}). Detonation kinetics is modeled using the Lee \& Tarver~\cite{LeeTarver1980} model. The constants for the kinetics model are taken from the paper~\cite{Urtiew2006}.

\begin{figure}[t]
	\centering
	\includegraphics[width=0.5\textwidth]{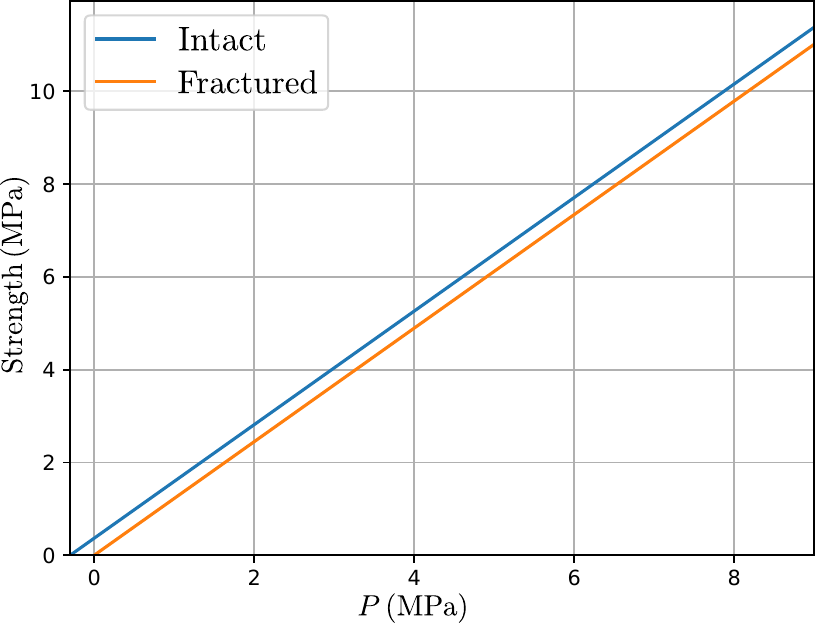}
	\caption{\label{fig:JHB_YP} Dependencies of $\sigma_{intact}(P)$ and $\sigma_{f}(P)$ for model JH-2}
\end{figure}

The JH-2~\cite{Johnson2008AnIC} brittle materials model is used to model the sand barrier:
$$\sigma_{intact} = \sigma_{intact}^0\frac{P + T_0}{P_{intact} + T_0},$$ 
$$\sigma_{f} = \sigma_{f}^0\frac{P}{P_{f}}.$$
The values of parameters used for sand modeling are given in Table~\ref{table:JH-2-sand}. The plots of $\sigma_{intact}(P)$ and $\sigma_{f}(P)$ are shown in Fig.~\ref{fig:JHB_YP}.

\begin{table}
	\center{
		\caption{\label{table:JH-2-sand} JH-2 model parameters for sand cylinder modeling.}
		\tabcolsep10pt
		\renewcommand{\arraystretch}{1.25}
		\begin{tabular}{|c|c|}
			\hline
			Parameter		& Value       \\
			\hline
			$\sigma_{intact}^0$, MPa	& $175.37$    \\
			\hline
			$P_{intact}$, MPa & $143$  \\
			\hline
			$\sigma_{f}^0$, MPa & $175$  \\
			\hline
			$P_{f}$, MPa & $143$  \\
			\hline
			Tensile strength $T_0$, MPa	& $0.3$    \\
			\hline
			Shear modulus $G$, MPa & 100 \\
			\hline
		\end{tabular}
		\tabcolsep15pt
	}
\end{table}

The Mie-Gr\"uneisen equation of state with shock Hugoniot as a reference is used as the equation of state for the sand:
$$P = P_{ref} + \gamma\rho(e - e_{ref}),$$
\begin{equation*}
	P_{ref} = \begin{cases}
		\rho_0 c_a^2\frac{1-x}{1-s_a(1-x)}, \, x\leq 1,
		\\
		\rho_0 c_a^2(1-x), \, x >1,
	\end{cases}
	e_{ref} = \begin{cases}
		\frac{1}{2}\left[\frac{c_a(1-x)}{1-s_a(1-x)}\right]^2, \, x\leq 1,
		\\
		c_a^2(1-x)^2/2, \, x >1,
	\end{cases}
	x = \frac{\rho_0}{\rho}.
\end{equation*}
Here $\rho_0 = 1674\,$ kg/m$^3$, $c_a = 1280\,$m/s, $\gamma = 2.17$, $s_a = 1.49$.

\subsection{Modeling results}
\begin{figure}[h]
	\includegraphics[width=\linewidth]{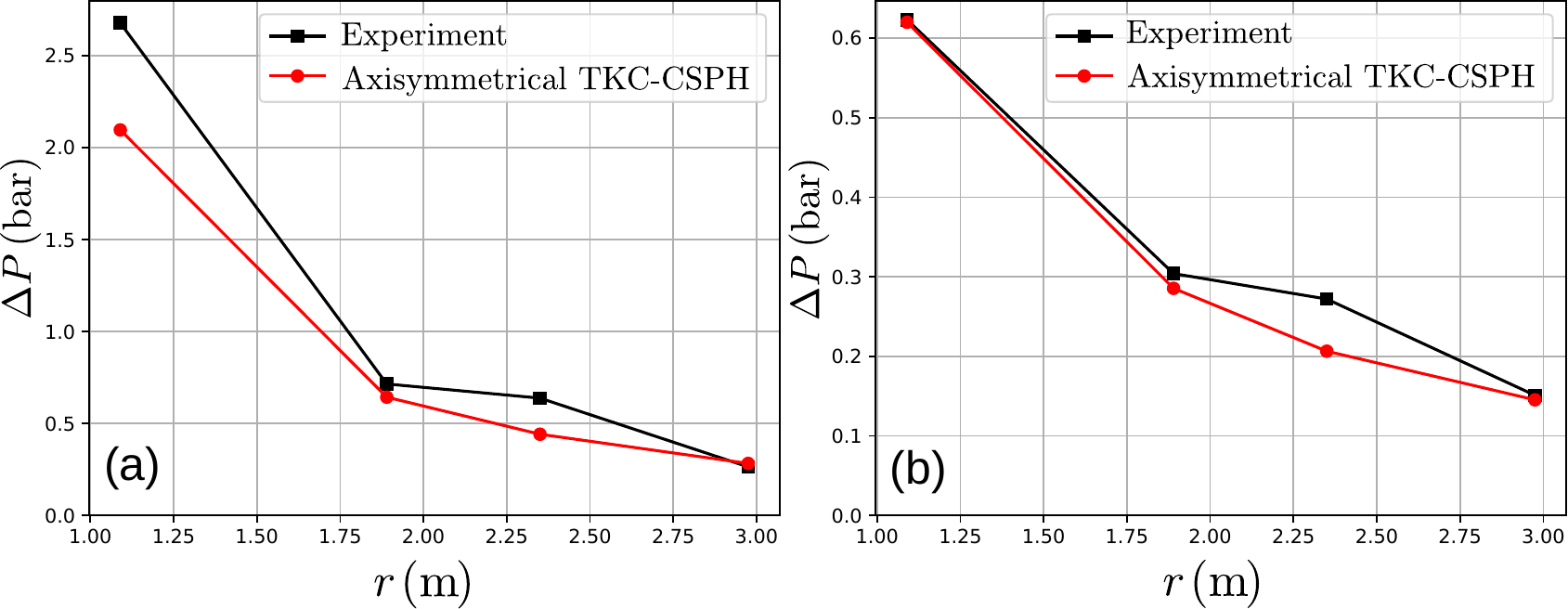}
	\caption{\label{fig:max-pressure-rise} Maximum pressure rise in the gauges relative to atmospheric pressure in the first experiment: (a) explosion in air, (b) explosion in a protective sand cylinder.}
\end{figure}

\begin{figure}[h]
	\includegraphics[width=\linewidth]{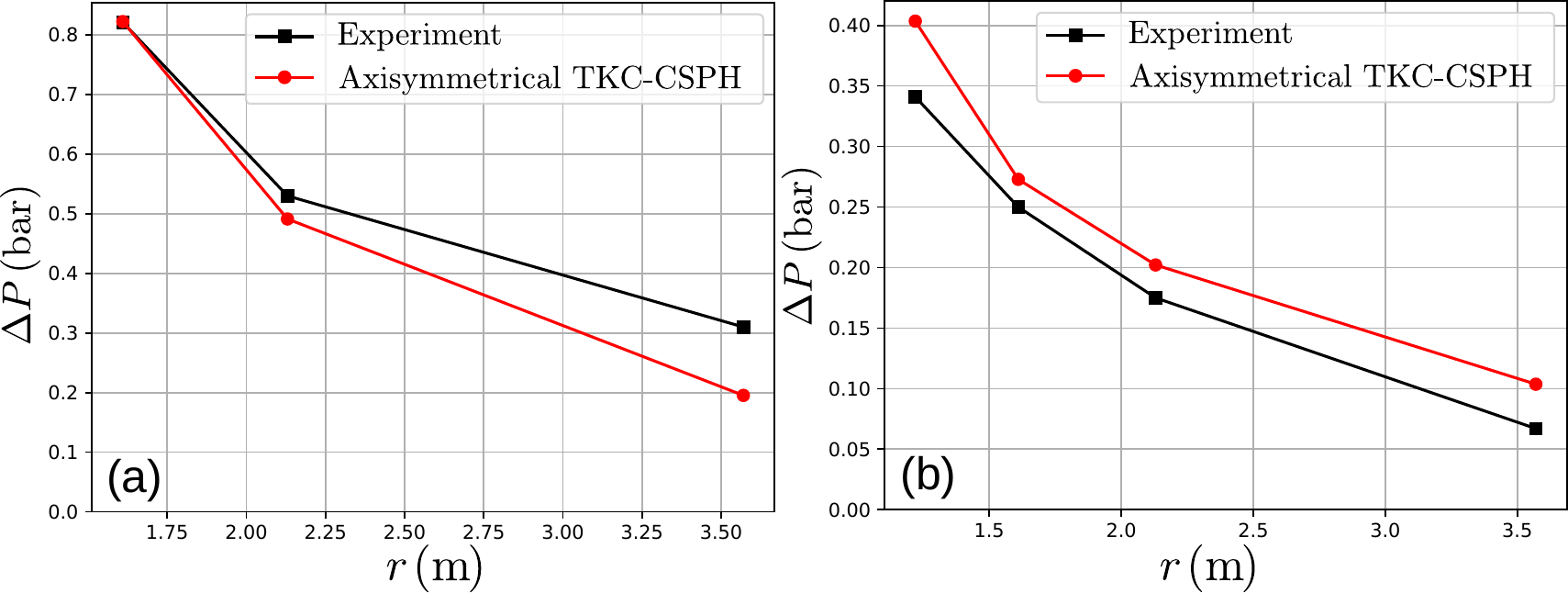}
	\caption{\label{fig:max-pressure-rise-variant-2} Maximum pressure rise in the sensors relative to atmospheric pressure in the second experiment: (a) explosion in air, (b) explosion in a protective sand cylinder.}
\end{figure}

\begin{table}
	\center{
		\caption{\label{table-gauges} Location of pressure gauges and simulation results.}
		\tabcolsep10pt
		\renewcommand{\arraystretch}{1.25}
		\begin{tabular}{c|c|c|c|c}
			\hline
			\hline
			\multicolumn{5}{c}{Experiment 1}\\
			\hline
			\multicolumn{3}{c|}{Gauge location} & $\Delta P,\,$bar (experiment)  & $\Delta P,\,$bar (SPH)\\
			Gauge number & $r,\,$m & Distance to the floor, m &   & \\
			\hline
			1 & 1.09 & 1.16 & 0.623 & 0.62\\
			\hline
			2 & 1.89 & 1.22 & 0.304 & 0.29\\
			\hline
			3 & 2.35 & 1.22 & 0.272 & 0.21\\
			\hline
			4 & 2.97 & 1.23 & 0.151 & 0.15\\
			\hline
			\hline
			\multicolumn{5}{c}{Experiment 2}\\
			\hline
			\multicolumn{3}{c|}{Gauge location} & $\Delta P,\,$bar (experiment)  & $\Delta P,\,$bar (SPH)\\
			Gauge number & $r,\,$m & Distance to the floor, m &   & \\
			\hline
			1 & 1.22 & 1.28 & 0.341 & 0.40\\
			\hline
			2 & 1.61 & 1.31 & 0.250 & 0.27\\
			\hline
			3 & 2.13 & 1.30 & 0.175 & 0.20\\
			\hline
			4 & 3.57 & 1.50 & 0.067 & 0.10\\
			\hline
			\hline
			\multicolumn{5}{c}{Explosion in air (without sand cylinder). The position of the charge is the same as in the first experiment. }\\
			\hline
			\multicolumn{3}{c|}{Gauge location} & $\Delta P,\,$bar (experiment)  & $\Delta P,\,$bar (SPH)\\
			Gauge number & $r,\,$m & Distance to the floor, m &   & \\
			\hline
			1 & 1.09 & 1.16 & 2.67 & 2.10\\
			\hline
			2 & 1.89 & 1.22 & 0.71 & 0.64\\
			\hline
			3 & 2.35 & 1.22 & 0.64 & 0.44\\
			\hline
			4 & 2.97 & 1.23 & 0.26 & 0.28\\
			\hline
			\hline
			\multicolumn{5}{c}{Explosion in air (without sand cylinder). The position of the charge is the same as in the second experiment. }\\
			\hline
			\multicolumn{3}{c|}{Gauge location} & $\Delta P,\,$bar (experiment)  & $\Delta P,\,$bar (SPH)\\
			Gauge number & $r,\,$m & Distance to the floor, m &   & \\
			\hline
			2 & 1.61 & 1.31 & 0.821 & 0.82\\
			\hline
			3 & 2.13 & 1.30 & 0.538 & 0.49\\
			\hline
			4 & 3.57 & 1.50 & 0.310 & 0.20\\
			\hline
			\hline
		\end{tabular}
		\tabcolsep15pt
	}
\end{table}

Fig.~\ref{fig:explosion2-in-sand} shows the evolution of sand barrier breakaway and shock wave propagation in the air surrounding the destroyed sand cylinder. A significant difference in the velocity between the expelling detonation products and fragments of the destroyed barrier is observed. The barrier during fracture takes away the part of the momentum from the detonation products, thus weakening the shock wave. At the same time, the barrier is fragmented into safe, low-velocity pieces.

Figs.~\ref{fig:max-pressure-rise} and \ref{fig:max-pressure-rise-variant-2} show the results of the explosion experiments with air and with a sand cylinder and the results of axisymmetric CSPH simulations with kernel gradient correction (TKC-CSPH). The maximum $\Delta P$ pressure rise in the gauges with respect to the atmospheric pressure $P_{atm}$, $\Delta P=P-P_{atm}$, is shown. The coordinates of the gauges are given in Table~\ref{table-gauges}.

\section*{Conclusions}
In this paper, a family of conservative schemes for the axisymmetric SPH contact method is proposed and considered. On the basis of Taylor series expansion by analogy with the work of \cite{Rublev2024} the corrected system of equations (TKC-CSPH) is obtained, and the correction is introduced to all equations of the method (continuity, motion and energy). The particle shifting technique~\cite{Michel2022} is used to maintain optimal particle packing. Particle splitting and merging are used to keep the particle size approximately the same in the computational domain.

As a condition near the axis of symmetry, it is proposed to interact particles near the axis of symmetry with particles mirrored beyond the axis of symmetry. It should be noted that schemes from the proposed family either do not allow interaction with reflected particles or lead to a significant decrease of accuracy due to large values of the $r/r_a$ ratio. Therefore, for about 3-4 layers of particles near the axis of symmetry, a local transition is carried out to the to the A.~N.~Parshikov scheme~\cite{Parshikov2000} (only for particles closer than twice smoothing length $h$ to the axis $z$ ($r < 2h$)). 

Compared with existing Riemann-based axisymmetric SPH methods, the advantage of the proposed approach is the complete elimination or at least significant suppression (depending on the presence or absence of particles near the axis of symmetry) of non-physical heating and acceleration of simulated objects.

Verification of the proposed approach is performed for the Riemann problem in an ideal gas (cylindrical Sod's test) and the Taylor bar test. Modeling of shock wave weakening in air by a breakaway sand barrier has been carried out. The modeling results are in a good agreement with the experimental data.

It is worth noting that the proposed approach has additional limitations due to the fact that particle splitting and merging techniques can lead to distortion of some small-scale flow structures and affect the conservation of total momentum and total energy during simulation.

\appendix
\section*{Appendix}
\subsection*{Kernel gradient correction for axisymmetric CSPH method}\label{A}
Let us analyze the approximation errors of the contact axisymmetric SPH method in the same way as it was done in~\cite{Rublev2024} for the Cartesian case. For this purpose, a Taylor series expansion of the expressions whose derivatives are evaluated using SPH sums is performed in the neighborhood of the point $\mathbf{r}_a$. The contact values are considered in the acoustic approximation (\ref{RazpadP}-\ref{RazpadU}).

After Taylor series expansion, the following expressions are obtained:
\begin{equation}\label{axisymmetric-csph-Teylor-continuity}
	\frac{d\varepsilon_a}{dt} = 
	\left(\boldsymbol{\nabla}\otimes\mathbf{U}_a\right) : \mathbf{L}_a  + \frac{U^r_a}{r_a} - \frac{1}{Z_a}\boldsymbol{\nabla} P_a \cdot \sum\limits_b v_br_{ab}\boldsymbol{\nabla}_aW_{ab} + O(h),
\end{equation}

\begin{multline}\label{axisymmetric-csph-Teylor-momentum}
	\frac{d\mathbf{U}_a}{dt} = 
	-\frac{2P_a}{\rho_a}\sum\limits_bv_b\boldsymbol{\nabla}_aW_{ab} - \frac{1}{\rho_a}\mathbf{L}_a\cdot\left(\boldsymbol{\nabla} P_a + \frac{P_a}{r_a}\mathbf{e}^r\right)  + \frac{P_a}{\rho_ar_a}\mathbf{e}^r + 
	\\
	\frac{Z_a}{\rho_a} \sum\limits_bv_b \left[\left(\boldsymbol{\nabla}\otimes\mathbf{U}_a\right):\left( \mathbf{e}^R_{ba}\otimes (\mathbf{r}_b - \mathbf{r}_a)\right)\right]\boldsymbol{\nabla}_a W_{ab} + O(h),
\end{multline}

\begin{multline}\label{axisymmetric-csph-Teylor-energy}
	\frac{dE_a}{dt} =
	-\frac{2P_a}{\rho_a}\mathbf{U}_a\cdot\sum\limits_bv_b\boldsymbol{\nabla}_aW_{ab}  + \frac{P_a}{\rho_aZ_a}\boldsymbol{\nabla} P_a\cdot\sum\limits_bv_br_{ab}\boldsymbol{\nabla}_aW_{ab}  -
	\\
	\frac{1}{\rho_a}\boldsymbol{\nabla} P_a\cdot\left(\mathbf{L}_a\mathbf{U}_a\right) - \frac{P_a}{\rho_a}\mathbf{L}_a:(\boldsymbol{\nabla}\otimes\mathbf{U}_a) - \frac{P_a}{\rho_ar_a}\mathbf{U}_a\cdot\left(\mathbf{L}_a\cdot\mathbf{e}^r\right) + 
	\\
	\frac{Z_a}{\rho_a}\sum\limits_bv_bU_a^R\left(\boldsymbol{\nabla}_a W_{ab}\otimes(\mathbf{r}_b - \mathbf{r}_a)\right):(\boldsymbol{\nabla}\otimes\mathbf{U}_a) + O(h),
\end{multline}
where $v_b = \frac{m_b}{2\pi\rho_br_b}$,
$$\mathbf{L}_a = \sum\limits_b v_b\boldsymbol{\nabla}_aW_{ab}\otimes (\mathbf{r}_b - \mathbf{r}_a).$$

Following the approach outlined in~\cite{Rublev2024} to introduce the kernel gradient correction into the equations of the CSPH method (Total Kernel Correction, TKC), one can write down the corrected equations of the ``harmonic mean'' scheme:
\begin{equation}\label{csph-normalized-continuity-axisymmetric}
	\frac{d\varepsilon_a}{dt} = -\sum\limits_{b}\frac{m_b}{2\pi\rho_b}\frac{r_a+r_b}{r_ar_b}(\overline{U}_{ab}^{*R} - \overline{U}_a^{R})\frac{W'_{ab}}{h_{ab}} + \frac{U^r_a}{ r_a},
\end{equation}
\begin{equation}\label{csph-normalized-motion-axisymmetric}
	\frac{d\mathbf{U}_a}{dt} = -\sum\limits_{b}\frac{m_b}{2\pi \rho_b\rho_a}\frac{r_a+r_b}{r_ar_b}\tilde{P}_{ab}^*\mathbf{L}_{ab}^{-1}\cdot\boldsymbol{\nabla}_aW_{ab} + \frac{P_a}{\rho_a r_a}\mathbf{e}^r.
\end{equation}
\begin{equation}\label{csph-normalized-energy-axisymmetric}
	\frac{dE_a}{dt} =
	\sum\limits_{b}\frac{m_b}{2\pi\rho_a\rho_b}\frac{r_a+r_b}{r_ar_b}\tilde{P}_{ab}^*\tilde{U}_{ab}^{*R} W'_{ab}.
\end{equation}
where $\overline{U}_a^{R} = \mathbf{U}_a\cdot \left(\mathbf{L}_{a}^{-1} \cdot\mathbf{e}^R_{ba}\right)$, $\tilde{U}_a^{R} = \mathbf{U}_a\cdot \left(\mathbf{L}_{ab}^{-1}\cdot \mathbf{e}^R_{ba}\right)$, $\mathbf{L}_{ab}^{-1} = \frac{1}{2}(\mathbf{L}_{a}^{-1} + \mathbf{L}_{b}^{-1})$.
$\overline{U}_{ab}^{*R}$, $\tilde{P}_{ab}^{*}$ and $\tilde{U}_{ab}^{*R}$ is calculated as
\begin{equation*}
	\overline{U}^{*R}_{ab} = \frac{\overline{U}_l^{R}Z_l + \overline{U}_r^{R}Z_r + (P_l - P_r)}{Z_l  + Z_r},
\end{equation*} 
\begin{equation*}
	\tilde{U}^{*R}_{ab} = \frac{\tilde{U}_l^{R}Z_l + \tilde{U}_r^{R}Z_r - P_r + P_l}{Z_l  + Z_r},
\end{equation*}
\begin{equation*}
	\tilde{P}^{*}_{ab} = \frac{P_rZ_l + P_lZ_r - Z_lZ_r\left( \tilde{U}_{r}^R - \tilde{U}_{l}^R \right)}{Z_l  + Z_r}.
\end{equation*}

\section*{Acknowledgments}
The author expresses his gratitude to A.N. Parshikov and S.A. Dyachkov for their helpful comments and suggestions.

\bibliography{library}
\end{document}